\documentclass[aps,prc,twocolumn]{revtex4-1}
\usepackage{amssymb}
\usepackage{amsmath}
\usepackage{wallpaper}
\usepackage{bm}
\usepackage[colorlinks,linkcolor=blue, urlcolor=blue, anchorcolor=blue, citecolor=blue]{hyperref}

\begin{document}

\title{Deconfinement  phase transition and quark condensate in compact stars}

\author{Hao-Miao~Jin$^{1,2,3}$}
\author{Cheng-Jun Xia$^{2,3}$}
\email{cjxia@yzu.edu.cn}
\author{Ting-Ting~Sun$^{1}$}
\email{ttsunphy@zzu.edu.cn}
\author{Guang-Xiong~Peng$^{4,5,6}$}
\email{gxpeng@ucas.ac.cn}

\affiliation{$^{1}${School of Physics and Microelectronics, Zhengzhou University, Zhengzhou 450001, China}
\\$^{2}${Center for Gravitation and Cosmology, College of Physical Science and Technology, Yangzhou University, Yangzhou 225009, China}
\\$^{3}${School of Information Science and Engineering, NingboTech University, Ningbo 315100, China}
\\$^{4}${School of Nuclear Science and Technology, University of Chinese Academy of Sciences, Beijing 100049, China}
\\$^{5}${Theoretical Physics Center for Science Facilities, Institute of High Energy Physics, P.O. Box 918, Beijing 100049, China}
\\$^{6}${Synergetic Innovation Center for Quantum Effects and Application, Hunan Normal University, Changsha 410081, China}}

\date{\today}

\begin{abstract}
We investigate systematically the possible deconfinement phase transition from nuclear matter to quark matter in compact stars. The properties of nuclear matter are fixed by expanding its binding energy to the order of $\rho^3$, while those of quark matter are predicted by an equivparticle model. The Maxwell construction is then applied for the quark-hadron mixed phase. By confronting compact star structures with pulsar observations, we obtain several EOSs that are compatible with the latest observations while supporting quark cores inside the most massive stars. It is found that the quark core is rather small and does not emerge for compact stars with $M\lesssim 2M_\odot$. The in-medium quark condensate of the stellar matter in those stars are then extracted within the framework of an equivparticle model, which decreases nonlinearly with density. At larger densities with pure quark matter, the quark condensate is still large and does not necessary decrease with density, indicating significant  nonperturbative contributions within the density regions covered by compact stars.
\end{abstract}

\maketitle

\section{\label{sec:intro}Introduction}
Exploring the equation of state (EOS) of dense matter has always been the common goal of astrophysics and nuclear physics. Using various messengers from isolated compact stars and their mergers, great progress has been made in achieving this goal especially since LIGO/VIRGO observed GW170817~\cite{LVC2018_PRL121-161101}. Among many interesting problems studied in the literature, for a long time, people have been committed to studying whether there is quark matter inside compact stars, the nature, place and time of hadron-quark phase transition, and if quark matter can be produced during their mergers, e.g., those in Refs.~\cite{Bauswein2019, Most2019}. Significant progresses have been made using various astrophysical observations such as the latest data from LIGO/VIRGO, NICER and Chandra observations~\cite{LVC2018_PRL121-161101, Riley2019_ApJ887-L21, Riley2021, Miller2019_ApJ887-L24, Miller2021}, where the properties of dense stellar matter can be constrained according to the most advanced theories and models~\cite{Li2019, Annala2020}. Nevertheless, ever since the early debate on whether the observational mass and radius of EXO 0748-676 can exclude the existence of a quark core in its center~\cite{Alford2007}, so far we have not reached consensus on the properties of dense stellar matter and the corresponding EOS.

One of the most interesting predictions in some dense matter theories is the possible phase transitions of nuclear matter into a singular state, including pion and kaon condensates as well as the deconfined quark matter~\cite{Weber2017}. It was shown that the kaon condensate may lead to a first-order phase transition at the onset of kaons~\cite{Maruyama1994_PLB337-19, Maruyama2006_PRC73-035802}. For the deconfinement phase transition, most of the effective models suggest it is of first-order~\cite{Glendenning1992, Peng2008_PRC77-065807, Li2015_PRC91-035803, Klahn2013_PRD88-085001, Bombaci2016_IJMPD-1730004}. In this case, the equilibrium phase transition from normal, low-density phase to pure singular phase occurs at a clearly defined pressure, accompanied by a density jump at the phase interface, i.e., the Maxwell construction with a bulk separation of the two phases.

By relaxing the conditions of local charge neutrality, it is possible that within a certain pressure range, two phases of dense matter coexist in the form of a mixture of low-density (nuclear) matter and high-density (quark) matter. Each phase is charged, and the mixture is only electrically neutral on average and fulfills the Gibbs condition~\cite{Glendenning1992,Glendenning1998,Glendenning1999}. The volume fraction $\chi$ occupied by the high-density phase increases from $\chi=0$ at the low-pressure boundary of the mixed phase to $\chi=1$ at the high-pressure boundary. If the surface tension of the interface between two phases is not too large, the mixed phase forming various types of geometrical structures is more favorable than those with a bulk separation between two phases~\cite{Heiselberg1993_PRL70-1355, Voskresensky2002_PLB541-93, Tatsumi2003_NPA718-359, Voskresensky2003_NPA723-291,Bejger2005, Endo2005_NPA749-333, Maruyama2007_PRD76-123015, Yasutake2014_PRC89-065803, Xia2019_PRD99-103017, Maslov2019_PRC100-025802, Xia2020_PRD102-023031}.

In this work we investigate systematically the possible phase transition from nuclear matter to quark matter in compact stars. For nuclear matter, we carry out a Taylor expansion of the binding energy to the order of $\rho^3$~\cite{Zhang2018_ApJ859-90}. For quark matter, we adopt an equivparticle model including both linear confinement and leading-order perturbative interactions~\cite{Peng2000_PRC62-025801, Wen2005_PRC72-015204, Xia2018_PRD98-034031, Xia2019_AIPCP2127-020029}.  Based on the obtained EOSs for both nuclear matter and quark matter, the properties of their mixed phase are then investigated adopting Maxwell construction, where the corresponding structures of hybrid stars are examined. We then obtain 1.8 million EOSs for hybrid star matter and examine the corresponding hybrid star structures. For those consistent with pulsar observations, it is found that the quark core is rather small and does not emerge for compact stars with $M\lesssim 2M_\odot$, which is consistent with the recent bayesian analysis adopting Nambu-Jona-Lasinio model for the quark phase~\cite{Pfaff2022}. We then extract the in-medium quark condensate in these dense stellar matter in the framework of an equivparticle model~\cite{Peng2002_PLB548-189}. It is found that the quark condensate is decreasing with density but rises in certain cases at the end of mixed phases, suggesting that the stellar matter in hybrid stars are highly nonperturbative even when a deconfinement phase transition takes place at $\rho\leq\rho_\mathrm{TOV}$, which is consistent with the recent studies assuming a smooth crossover from nuclear matter to quark matter~\cite{Minamikawa2021,Jin2022}.

This paper is organized as follows. The theoretical framework in obtaining the EOSs of nuclear matter, quark matter, and their mixed phase are presented in Section~\ref{sec:the}, while the equivparticle model for extracting the corresponding in-medium quark condensate is introduced as well. In Section~\ref{sec:num}, the numerical results on the constrained properties of hybrid star matter are presented, in which the quark condensates are extracted. Finally, we give a summary in Section~\ref{sec:con}.

\section{\label{sec:the}Theoretical framework}
\subsection{\label{sec:the_NS}Nuclear matter}
To obtain the EOSs of nuclear matter, we first adopt the Taylor expansion method for the binding energy $\epsilon_{0}(\rho)$ in symmetric nuclear matter (SNM) and the symmetry energy $S(\rho)$ with the baryon number density $\rho=\rho_n + \rho_p$. The binding energy per nucleon for nuclear matter in isospin asymmetry $\delta =(\rho_n - \rho_p)/\rho$ reads
\begin{equation}
\epsilon(\rho, \delta)\approx \epsilon_{0}(\rho)+S(\rho)\delta ^{2}. \label{Eq:Et}
\end{equation}
Omitting higher order terms, we then have
\begin{eqnarray}
\epsilon_{0} &=&  \epsilon_0(\rho_0)+\frac{K}{18} x^{2}+\frac{J}{162} x^{3}, \label{Eq:E0}\\
S     &=&   S(\rho_0) +\frac{L}{3} x +\frac{K_\mathrm{sym}}{18}x^{2}  + \frac{J_\mathrm{sym}}{162}x^{3}, \label{Eq:Es}
\end{eqnarray}
with $x\equiv\left({\rho}/{\rho_0} - 1\right)$, the binding energy $\epsilon_{0}(\rho _{0}) \approx -15.9$ MeV and the symmetry energy $S(\rho_{0}) = 31.7 \pm 3.2$ MeV at the saturation density $\rho_{0} = 0.16$ fm$^{-3}$. In Eqs.~(\ref{Eq:E0}) and (\ref{Eq:Es}), $K$ and $J$ are the incompressibility and skewness of SNM, while $L$, $K_\mathrm{sym}$, and $J_\mathrm{sym}$ are the slope, curvature, and skewness of the symmetry energy, which are all fixed at $\rho=\rho_{0}$. According to extensive nuclear and astrophysical studies, they are constrained with $K = 240 \pm 20$ MeV~\cite{Shlomo2006_EPJA30-23}, $L = 58.7 \pm 28.1$ MeV~\cite{Li2013_PLB727-276, Oertel2017_RMP89-015007, Zhang2020_PRC101-034303, Essick2021_PRL127-192701} and $K_\mathrm{sym}= -107 \pm 88$ MeV~\cite{Li2021_Universe7-182}. Meanwhile, the symmetry energy at $\rho_\mathrm{on}=0.1\ \mathrm{fm}^{-3}$ is well constrained with $S(\rho_\mathrm{on}) =25.5 \pm 1.0$ MeV~\cite{Centelles2009_PRL102-122502, Brown2013_PRL111-232502}, indicating  a relation between the symmetry energy and its slope at $\rho=\rho_{0}$, i.e., $S(\rho_{0}) \approx 26 + L/9$ MeV~\cite{Horowitz2001_PRL86-5647}. Note that expansion in Eqs.~(\ref{Eq:E0}) and (\ref{Eq:Es}) will not converge at supersaturated density, which can be fixed by other expansion techniques~\cite{Margueron2018_PRC97-025805, Cai2021_PRC103-054611}.

For given coefficients in Eqs.~(\ref{Eq:E0}) and (\ref{Eq:Es}), the energy density of nuclear matter is written as
\begin{equation}
E_\mathrm{b}(\rho, \delta)=\rho \epsilon(\rho, \delta) +\rho M_{N}, \label{Eq:E_expand}
\end{equation}
where the binding energy $\epsilon(\rho, \delta)$ is fixed by Eq.~(\ref{Eq:Et}) and $M_{N}=938$ MeV is the rest mass of nucleons.
By further including the contributions of leptons, the energy density $E^{N}$ of $npe\mu$ nuclear matter reads
\begin{equation}
  E^{N} =E_\mathrm{b}(\rho, \delta) + \sum_{l=e,\mu}E_{l}(\rho_{l},m_{l}).
 \label{Eq:EH}
\end{equation}
Here $E_l (\rho_{l},m_{l})$ is the energy density of leptons, which is determined by
\begin{equation}
E_{l}(\rho_{l},m_{l})=\frac{m_l^{4}}{8\pi ^2}f\left(\frac{\sqrt[3]{3\pi^2 \rho_l}}{m_l}\right),
 \end{equation}
with the electron mass $m_e=0.511$ MeV, muon mass $m_\mu=105.66$ MeV, and
\begin{equation}
f(x)\equiv \left[x\sqrt{1+x^2}\left(1+2x^2\right)-\mathrm{arcsh}(x)\right]. \label{Eq:fx}
\end{equation}

The chemical potential of particle type $i$ can then be calculated from
\begin{equation}
\mu_i=\frac{\partial E^{N}}{\partial \rho_i}.
 \label{Eq:MU}
\end{equation}
According to basic thermodynamic relations, the pressure is obtained with
\begin{equation}
  P^{N} = -E^{N}+ \sum_{i}\mu_{i} \rho_{i}.
  \label{Eq:PH}
 \end{equation}
Through the $\beta$-equilibrium condition $\mu _{e}=\mu_{\mu }=\mu_{n}-\mu_{p}$ and the charge neutrality condition $n_\mathrm{ch}^{N}=\rho _{p}-\rho _{e}-\rho_\mu=0$, we can obtain the isospin asymmetry $\delta(\rho)$ and relative particle fractions ($\rho_i/\rho$ with $i=p,n,e,\mu$) of nuclear matter in compact stars at different densities.

\subsection{\label{sec:the_QM}Quark matter}
Here we assume that quark matter is comprised of up ($u$) and down ($d$) quarks, while charge neutrality is maintained by including electrons ($e$) and muons ($\mu$). Note that we have neglected strangeness for simplicity. At zero temperature, the energy density of a free quark system is
\begin{equation}
E^{Q} = \sum_{i}\frac{m_{i}^{4}g_{i}}{16\pi^{2}}  f\left(\frac{\nu_i}{m_i}\right), \label{Eq:EQ}
\end{equation}
where $f(x)$ is fixed by Eq.~(\ref{Eq:fx}), $g_{i}$ ($g_{e,\mu}=2, g_{u,d}=6$) the degeneracy factor of quarks, and $\nu_{i}$ ($i=u,d,e,\mu$) the particle's Fermi momentum. It is connected to the particle number density by
\begin{equation}
\rho_{i}=\frac{g_{i}\nu_{i}^{3}}{6\pi^{2}},
\label{Eq:nu}
\end{equation}
with the corresponding chemical potential
\begin{equation}
\mu_{i}^{*}=\sqrt{\nu_{i}^{2}+m_{i}^{2}}. \label{Eq:mueff}
\end{equation}

In the framework of equivparticle models~\cite{Peng1999_PRC61-015201, Peng2002_PLB548-189, Wen2005_PRC72-015204, Xia2014_PRD89-105027}, quarks are treated as quasifree particles with density-dependent equivalent masses, while the energy density takes the same form as Eq.~(\ref{Eq:EQ}). Taking into account both the linear confinement and leading-order perturbative interactions, the quark mass scaling is given by~\cite{Xia2014_PRD89-105027}
\begin{equation}
 m_{i}(\rho)=m_{i0}+m_\mathrm{I}=m_{i0}+\frac{D}{\rho^{1/3}}+C \rho^{1/3}.
 \label{Eq:md}
\end{equation}
Here we assume an exact isospin symmetry with $m_{u0}=m_{d0}=m_0=3.45$ MeV being the average current mass of $u$ and $d$ quarks. The confinement parameter $D$ is connected to the string tension, and the perturbative strength parameter $C$ is linked to the strong coupling constant. According to previous investigations, the parameter $\sqrt{D}$ lies in the range (147, 270) MeV~\cite{Wen2005_PRC72-015204} and $C\lesssim$ 1.2~\cite{Xia2014_PRD89-105027}. In this work, we examine all possible combinations by varying those parameters within $125\ \mathrm{MeV}\leq \sqrt{D} \leq 270$ MeV and $-1\leq C \leq 1$.

Due to the density dependence of quark masses, the real chemical potential have an additional term and $\mu_{i}^{*}$ in Eq.~(\ref{Eq:mueff}) should be viewed as an effective one. We write the real chemical potential as
\begin{equation}
\mu_{i}=\frac{\partial E^{Q}}{\partial \rho_i}=\mu_{i}^{*}+\mu_\mathrm{I}=\mu_{i}^{*}+\frac{1}{3}\frac{\partial m_{i}(\rho)}{\partial \rho}\frac{\partial E^{Q}}{\partial m_\mathrm{I}}.
\label{Eq:MI}
\end{equation}
Therefore, the pressure of quark matter is obtained by
\begin{equation}
P^{Q}=-E^{Q}+\sum_{i}\mu_{i}\rho_{i}.
\label{Eq:PQ}
\end{equation}

Due to the weak interactions, the chemical potentials $\mu_{i}$ ($i=u,d,e,\mu$) satisfy the weak equilibrium condition:
\begin{equation}
\mu_{d}-\mu_{u}=\mu_{e}=\mu_{\mu}.
\label{Eq:EC}
\end{equation}
Additionally, the quark matter in compact stars should fulfill the conditions of charge neutrality 
\begin{equation}
n_\mathrm{ch}^{Q}=\frac{2}{3}\rho_{u}-\frac{1}{3}\rho_{d}-\rho_{e}-\rho_{\mu}=0
\label{Eq:NC}
\end{equation}
and baryon number conservation
\begin{equation}
\frac{1}{3}(\rho_{u}+\rho_{d})=\rho.
\label{Eq:BN}
\end{equation}
Therefore, Eqs.~(\ref{Eq:EC})--(\ref{Eq:BN}) are four equations about the four chemical potentials $\mu_{i}$ and can be solved at a given baryon number density $\rho$.

According to Eq.~(\ref{Eq:MI}), the real and effective chemical potentials for each flavor of quarks differ merely by a common quantity $\mu_\mathrm{I}$. Thus the effective chemical potentials also satisfy the similar weak equilibrium conditions
\begin{equation}
\mu_{d}^{*}-\mu_{u}^{*}=\mu_{e}=\mu_{\mu}.
\label{Eq:WEC}
\end{equation}
Because electrons and  muons do not participate in strong interactions, the corresponding mass is constant. Consequently, the real and effective chemical potentials of electrons and  muons are the same.

\subsection{\label{sec:the_MP}Quark-hadron mixed phase}
\begin{figure}[!ht]
\centering
 \includegraphics[width=\linewidth]{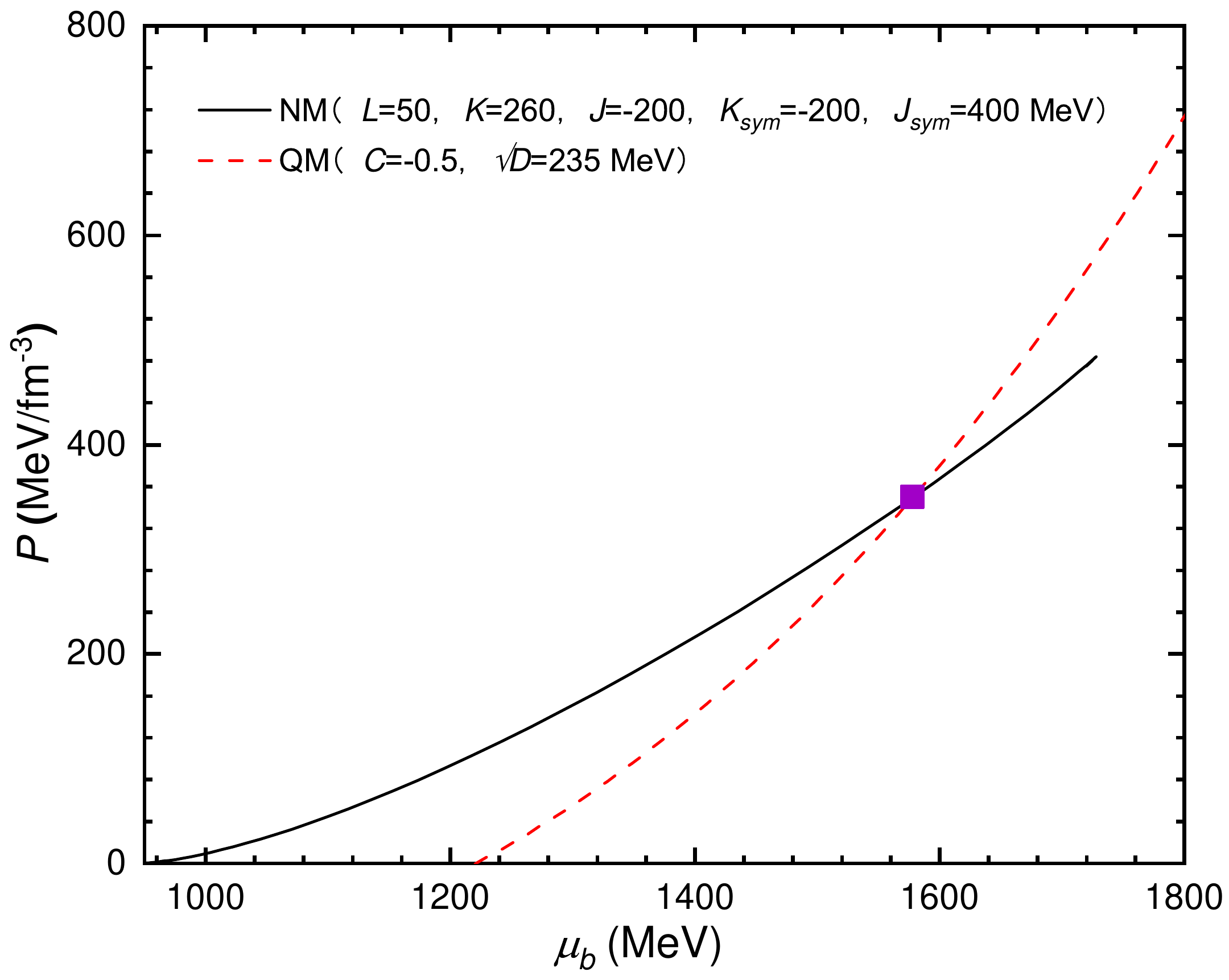}
 \caption{\label{Fig:NQeos}Pressure $P$ as a function of baryon chemical potential $\mu_\mathrm{b}$ for nuclear matter (NM) and quark matter (QM).}
\end{figure}

To investigate the properties of mixed phase and its implication on compact star structures, we adopt the Maxwell construction, which should be valid if the surface tension exceeds a critical value~\cite{Xia2019_PRD99-103017}. In this case, at a given baryon chemical potential $\mu_\mathrm{b}$, the dynamic stability condition needs to be satisfied $P^{N}(\mu_\mathrm{b})=P^{Q}(\mu_\mathrm{b})$. A simple example is illustrated in Fig.~\ref{Fig:NQeos}, where the pressures of nuclear matter and quark matter intersect (i.e. purple square), indicating a phase transition that forms the mixed phase.

For the Maxwell construction, the size of mixed phase is much larger than the Debye screening length $\lambda_{D}$, so that the Coulomb repulsion effectively restores the local charge neutrality condition~\cite{Xie2021}. Thus, we have
\begin{equation}
\mu_{e}^{N}\neq \mu_{e}^{Q},\qquad  n_\mathrm{ch}^{N}=0, \qquad n_\mathrm{ch}^{Q}=0.
\label{Eq:nch}
\end{equation}
Here $n_\mathrm{ch}^{I}$ is the charge density of phase $I (I=N,Q)$.

To obtain the properties of mixed phase, we write the total energy density~\cite{Ang2009} as
\begin{equation}
E_\mathrm{t}=(1-\chi)E^{N}+\chi E^{Q}.
\label{Eq:ET}
\end{equation}
Here the quark fraction $\chi \equiv V^{Q}/V$ ($0\leq \chi \leq 1$)  with $V^{Q}$ being the volume occupied by quarks and $V$ the total volume.
The structure of mixed phase can be fixed by minimizing the energy density in Eq.~(\ref{Eq:ET}) at a given total baryon number density $\rho=(1-\chi)\rho^{N}+\chi \rho^{Q}$, which indicates \begin{equation}
\mu_\mathrm{b}=\mu_\mathrm{b}^{N}= \mu_\mathrm{b}^{Q}, \ \ P^{N}(\mu_\mathrm{b})=P^{Q}(\mu_\mathrm{b})
\label{Eq:PT}
\end{equation}
with $\mu_\mathrm{b}^{N}=\mu_n$ and $\mu_\mathrm{b}^{Q}=\mu_u+2\mu_d$. Then the properties of the mixed phase can be obtained.

\subsection{\label{sec:the_Equiv}Equivparticle model on quark condensate}
The Hamiltonian density for any QCD system can be written as
\begin{equation}
 H_\mathrm{QCD}=H_{K}+\sum_{i}m_{0i}\bar{q_{i}}q_{i} + H_\mathrm{I}, \label{Eq:HQCD}
\end{equation}
where $H_{K}$ represents the kinetic term, $m_{i0}$ the quark current mass, and $H_\mathrm{I}$ the interaction term. Meanwhile, in the framework of equivparticle model, we can define an equivalent Hamiltonian density with a variable quark mass as
\begin{equation}
 H_\mathrm{eqv}=H_{K}+\sum_{i}m_{i}\bar{q_{i}}q_{i}, \label{Eq:Hequiv}
\end{equation}
where $m_{i}$ is the equivalent mass of quark $i$. The interaction part $H_\mathrm{I}$ in Eq.~(\ref{Eq:HQCD}) is then included in the equivalent mass $m_{i}$, while $m_{i}$ can be divided into two parts, i.e.,
\begin{equation}
 m_{i}=m_{i0}+m_\mathrm{I} \label{Eq:mass}
\end{equation}
with $m_\mathrm{I}$ accounts for the strong interaction among quarks. In order for $H_\mathrm{eqv}$ to reflect the characteristics of the original QCD system, we then obtain
\begin{equation}
\langle H_\mathrm{eqv} \rangle_\rho - \langle H_\mathrm{eqv} \rangle_0 = \langle H_\mathrm{QCD} \rangle_\rho - \langle H_\mathrm{QCD} \rangle_0. \label{Eq:Heq}
\end{equation}
Here $\langle O \rangle_\rho \equiv \langle\rho | O |\rho \rangle$ represents the expectation value of operator $O$ in dense stellar matter with density $\rho$, while the vacuum contribution $\langle O \rangle_0 \equiv \langle 0 | O | 0 \rangle$ needs to be subtracted.

Substituting Eqs.~(\ref{Eq:HQCD}) and (\ref{Eq:Hequiv}) into Eq.~(\ref{Eq:Heq}), we then obtain the interacting part of the equivalent mass
\begin{equation}
  m_\mathrm{I}=\frac{E_\mathrm{I}}{\Sigma_{i}(\langle\bar{q_{i}}q_{i}\rangle_\rho - \langle\bar{q_{i}}q_{i}\rangle_{0})}, \label{Eq:massI}
\end{equation}
with the interacting energy density
\begin{equation}
 E_\mathrm{I}=\langle H_\mathrm{I} \rangle_\rho -\langle H_\mathrm{I}\rangle_0.
\end{equation}
Note that in obtaining Eq.~(\ref{Eq:massI}) we have assumed that the quark condensate is uniformly distributed, i.e., by taking the spatial average. In such cases, any local fluctuations in the nucleon or mixed phase are ignored, corresponding to $\langle m_{i}\bar{q_{i}}q_{i}\rangle = m_{i}\langle\bar{q_{i}}q_{i}\rangle$.

We have neglected the contribution of strange quarks throughout the density range of compact stars, where only  two quark flavors ($u$ and $d$) are considered. By assuming an exact isospin symmetry, we have $m_{u0}=m_{d0}\equiv m_{0}$, $m_{u}=m_{d}\equiv m$, $\langle\bar{u}u \rangle_0 = \langle\bar{d}d \rangle_0 \equiv \langle\bar{q}q\rangle_\rho$, and $\langle\bar{u}u \rangle_\rho + \langle\bar{d}d \rangle_\rho \equiv 2 \langle\bar{q}q\rangle_\rho$. The in-medium quark condensate can then be obtained by rewriting Eq.~(\ref{Eq:massI}), which gives
\begin{equation}
 \frac{\langle\bar{q}q\rangle_\rho} {\langle\bar{q}q\rangle_0}=1-\frac{1}{3n^*}\frac{E_\mathrm{I}}{m_\mathrm{I}}. \label{Eq:cond}
\end{equation}
According to the Gellman-Oakes-Renner relation $-2m_{0}\langle\bar{q}q\rangle_{0}=m_{\pi}^{2}f_{\pi}^{2}$~\cite{Gell-Mann1968_PR175-2195}, the vacuum quark condensate $\langle\bar{q}q\rangle_0$ and chiral recovery density $n^*$ in the model independent linear expression~\cite{Cohen1992_PRC45-1881} can be obtained with
\begin{equation}
   n^{*}=-\frac{2}{3}\langle\bar{q}q\rangle_{0}=\frac{m_{\pi}^{2}f_{\pi}^{2}}{3m_0} = 2.141\ \mathrm{fm}^{-3}
   \label{Eq:gr}
\end{equation}
where $m_{\pi}\approx 140$ MeV, $f_{\pi}\approx 93.2$ MeV and $m_0=3.45$ MeV are the pion mass, the pion decay constant and the average current mass of light quarks, respectively.

Since the $H_\mathrm{eqv}$ has the same form of free system with the equivalent quark mass  $m_{i}$, the energy density can be obtained by
\begin{equation}
E(\rho_u, \rho_d, m) = \frac{m^{4}g}{16\pi^{2}} \sum_{q=u,d} f\left(\frac{\sqrt[3]{6\rho_q\pi^2}}{m\sqrt[3]{g}}\right). \label{Eq:E_equiv}
\end{equation}
Here $f(x)$ is given by Eq.~(\ref{Eq:fx}) and $g = 2 ($spin$) \times 3 ($color$) =6$. To fix the equivalent quark mass $m$ for nuclear matter, quark matter, and their mixed phase, we need to reproduce the energy density with Eq.~(\ref{Eq:E_equiv}) by varying $m$, i.e.,
\begin{equation}
  E(\rho_u, \rho_d, m) =
 \left\{\begin{array}{l}
   E_\mathrm{b}(\rho, \delta),\  \text{nuclear matter}\\
   E^{Q}-E_e-E_\mu, \ \text{quark matter}\\
   E_\mathrm{t}-E_e-E_\mu, \ \  \text{mixed phase}\\
 \end{array}\right., \label{Eq:meqv_cal}
\end{equation}
where $E_\mathrm{b}$, $E^{Q}$, and $E_\mathrm{t}$ are fixed by Eqs.~(\ref{Eq:E_expand}), (\ref{Eq:EQ}), and (\ref{Eq:ET}), respectively. The quark densities take the same values as the average densities of nuclear matter ($\chi=0$), quark matter ($\chi=1$), and their mixed phase ($0<\chi<1$), where $\rho_{u,d}=(1-\chi)\rho_{u,d}^{N}+\chi \rho_{u,d}^{Q}$ with $\rho_u^{N} = 2\rho_p + \rho_n = (3-\delta)\rho/2$ and $\rho_d^{N} = \rho_p + 2\rho_n= (3+\delta)\rho/2$. Meanwhile, the leptons in dense stellar matter have nothing to do with strong interactions, whose energy density is obtained with $E_{e,\mu}=(1-\chi)E_{e,\mu}^{N}+\chi E_{e,\mu}^{Q}$ and is subtracted in Eq.~(\ref{Eq:meqv_cal}).  Once we fix the equivalent mass $m$, the interacting energy density of dense stellar matter can be obtained by subtracting the energy density of free quarks, i.e.,
\begin{equation}
  E_\mathrm{I}= E(\rho_u, \rho_d, m) - E(\rho_u, \rho_d, m_0). \label{Eq:EI}
\end{equation}
The interacting part of the equivalent mass can then be obtained with
\begin{equation}
  m_\mathrm{I}= m-m_{0}.  \label{Eq:mI_cal}
\end{equation}
Note that for the deconfined quark matter, the interacting part of the equivalent mass can be obtained with Eq.~(\ref{Eq:md}), i.e.,
\begin{equation}
 m_\mathrm{I}^{Q}=\frac{D}{\rho^{1/3}}+C \rho^{1/3}.
 \label{Eq:mc}
\end{equation}
Based on the obtained values for $m_\mathrm{I}$ and $E_\mathrm{I}$, the in-medium quark condensate in dense stellar matter is then calculated by Eq.~(\ref{Eq:cond}).

\section{\label{sec:num}Results and discussions}
In order to constrain the properties of dense stellar matter, as was done in Ref.~\cite{Zhang2018_ApJ859-90}, we carry out extensive calculations to obtain the EOSs according to the formulae introduced in Secs.~\ref{sec:the_NS}--\ref{sec:the_MP}. Note that we have varied the parameters for nuclear matter in the steps of 10, 20, 200, 100 and 200 MeV within the range of $L=40\sim80$ MeV, $K=220\sim260$ MeV, $J=-800\sim400$ MeV, $K_\mathrm{sym} = -400\sim100$ MeV and $J_\mathrm{sym} = -200\sim800$ MeV, and for quark matter in the steps of 5 and 0.1 within the range of $\sqrt{D}=125 \sim 270$ MeV and $C=-1\sim 1$ MeV, respectively. We then have totally 1,810,620 EOSs for hybrid star matter.

\begin{table*}
\caption{\label{tableb} The saturation properties of nuclear matter in Eqs.~(\ref{Eq:E0}-\ref{Eq:Es}) and the parameter sets of Eq.~(\ref{Eq:md}) for quark matter, which are constrained according to pulsar observations, i.e., the tidal deformability $70\leq \Lambda_{1.4}\leq 580$ (90\% credible region)~\cite{Abbott2019}, the radii $R_{1.4}=12.45\pm 1.30$ km and $R_{2.08}=12.35\pm 1.50$ km (95\% credible region)~\cite{Miller2021} with the maximum mass $M_\mathrm{TOV} \geq 2.08 M_{\odot}$~\cite{Cromartie2020_NA4-72}.}
\centering
\begin{tabular}{c|cccccccc|ccccc|c}
  \hline  \hline
  & $S(\rho_{0})$ & $L$ & $K$ & $J$ & $K_\mathrm{sym}$& $J_\mathrm{sym}$  &$\sqrt{D}$ & $C$ &$R_{1.4}$ & $\Lambda_{1.4}$ &$R_{2.08}$& $\rho_\mathrm{TOV}$ & $M_\mathrm{TOV}$ & $v_\mathrm{max}$ \\
 & MeV &   MeV        &  MeV & MeV & MeV      &    MeV     & MeV & & km &       &  km      &    fm$^{-3}$ &  $M_\odot$ &  $c$ \\  \hline
\uppercase\expandafter{\romannumeral1} & 33.8 & 70 & 260 &    $-200$   & $-400$ & 800 &265  &$-0.9$&  11.77 & 286 & 11.28 & 1.37 & 2.09 & 0.99 \\
\uppercase\expandafter{\romannumeral2} & 33.8 & 70 & 260 &    $-200$   & $-400$ & 800 &235  &$-0.5$&  11.78 & 289 & 11.22 & 1.35 & 2.13 & 0.99 \\
\uppercase\expandafter{\romannumeral3} & 32.7 & 60 & 260 &    $-200$   & $-300$ & 600 &260  &$-0.8$&  11.77 & 316 & 11.21 & 1.40 & 2.09 & 0.92 \\
\uppercase\expandafter{\romannumeral4} & 31.6 & 50 & 260 &    $-200$   & $-200$ & 400 &245  &$-0.6$&  11.82 & 344 & 11.14 & 1.40 & 2.08 & 0.85 \\
\uppercase\expandafter{\romannumeral5} & 31.6 & 50 & 260 &    $-200$   & $-200$ & 400 &235  &$-0.5$&  11.82 & 344 & 11.14 & 1.36 & 2.08 & 0.85 \\
\uppercase\expandafter{\romannumeral6} & 30.4 & 40 & 260 &    $-200$   & $-200$ & 600 &165  &$ 0.1$&  11.86 & 348 & 11.59 & 0.97 & 2.12 & 0.83 \\
\uppercase\expandafter{\romannumeral7} & 30.4 & 40 & 260 &    $-200$   & $-200$ & 600 &155  &$ 0.2$&  11.86 & 347 & 11.56 & 0.96 & 2.14 & 0.83 \\
 \hline
\end{tabular}
\end{table*}

By analyzing the gravitational waves emitted from the binary neutron star merger event GW170817, the tidal deformability of $1.4 M_{\odot}$ neutron star are constrained within $70\leq \Lambda_{1.4}\leq 580$, corresponding to the radii $11.9^{+1.4}_{-1.4}$ km~\cite{LVC2018_PRL121-161101}. Based on pulse-profile modeling~\cite{Watts2019_SCPMA62-29503}, both the mass and radius of a pulsar can be measured, e.g., the observations of PSR J0030+0451 and PSR J0740+6620 have placed the radii of $1.4 M_{\odot}$ and $2.08 M_{\odot}$ compact stars at $12.45\pm 0.65$ km and $12.35\pm 0.75$ km (68\% credible region)~\cite{Miller2021}. Combining all those observations, the most stringent constraint on the EOSs of dense matter can be obtained~\cite{Li2020, Zhang2020_PRC101-034303}.
More specifically, we constrain the properties of dense stellar matter at $\rho\leq\rho_\mathrm{TOV}$ according to the tidal deformability $70\leq \Lambda_{1.4}\leq 580$~\cite{LVC2018_PRL121-161101}, the radii $R_{1.4}=12.45\pm 1.30$ km and $R_{2.08}=12.35\pm 1.50$ km~\cite{Miller2021}, where the maximum mass $M_\mathrm{TOV} \geq 2.08 M_{\odot}$~\cite{Fonseca2021} with $\rho_\mathrm{TOV}$ being the corresponding central density. In Table~\ref{tableb} we present the parameter sets that meet the constraints, as well as the corresponding hybrid star properties and maximum sound velocity for hybrid star matter. For those consistent with observations and support a quark core inside a compact star, we have constrained the parameters $K\approx260$ MeV, $J \approx -200$ MeV, $L\approx40\sim 70$ MeV, $K_\mathrm{sym}\approx -400\sim-200$ MeV, and $J_\mathrm{sym} \approx 400\sim800$ MeV for nuclear matter and $\sqrt{D}=155 \sim 265$ MeV and $C=-0.9\sim 0.1$ MeV for the equivparticle model of quark matter. Note that we have disregarded the cases where the first-order deconfinement phase transition takes place at $\rho>\rho_\mathrm{TOV}$, which were examined in our previous study~\cite{Jin2022}.

\begin{figure*}[!ht]
  \centering
  \includegraphics[width=0.47\linewidth]{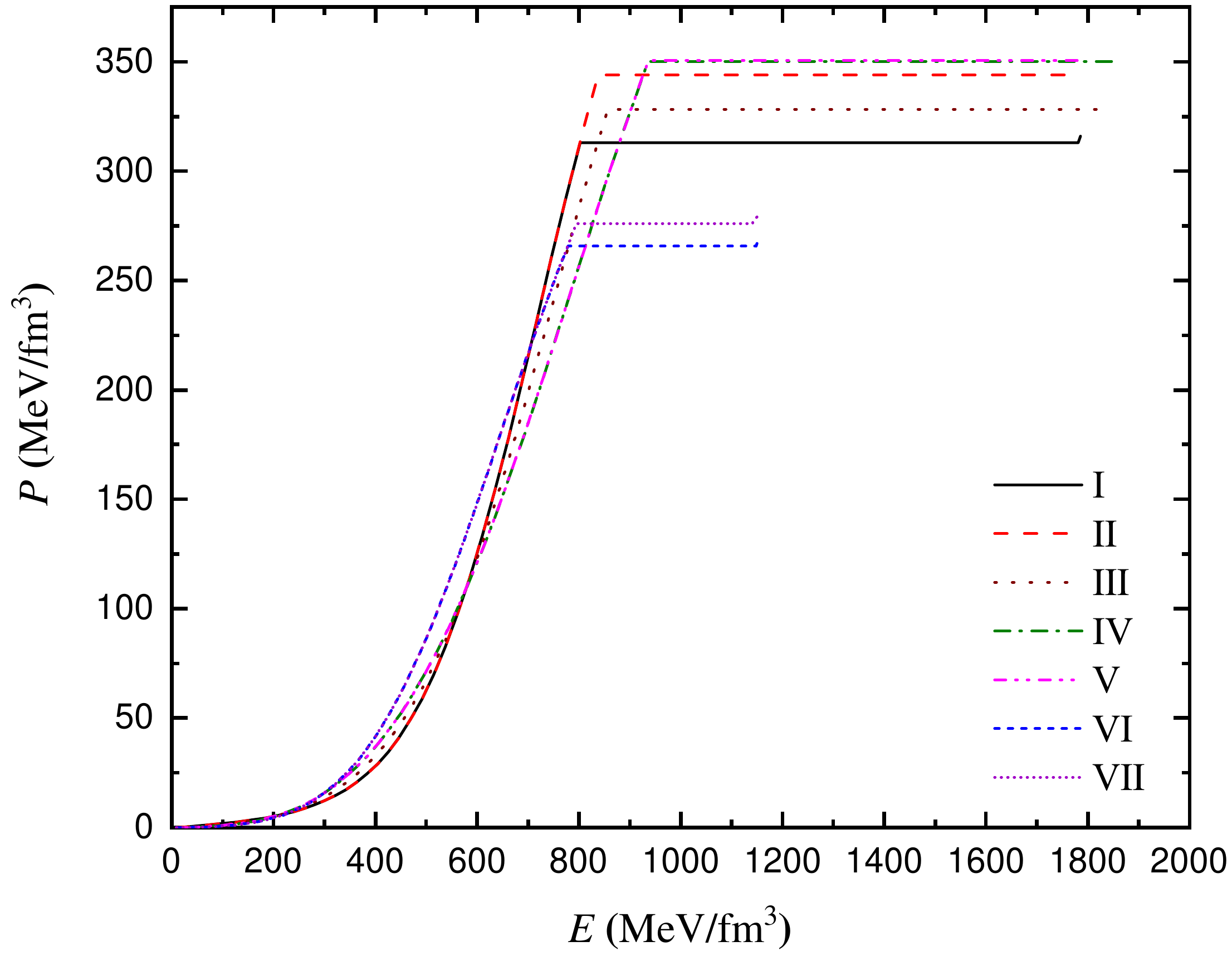}
  \includegraphics[width=0.48\linewidth]{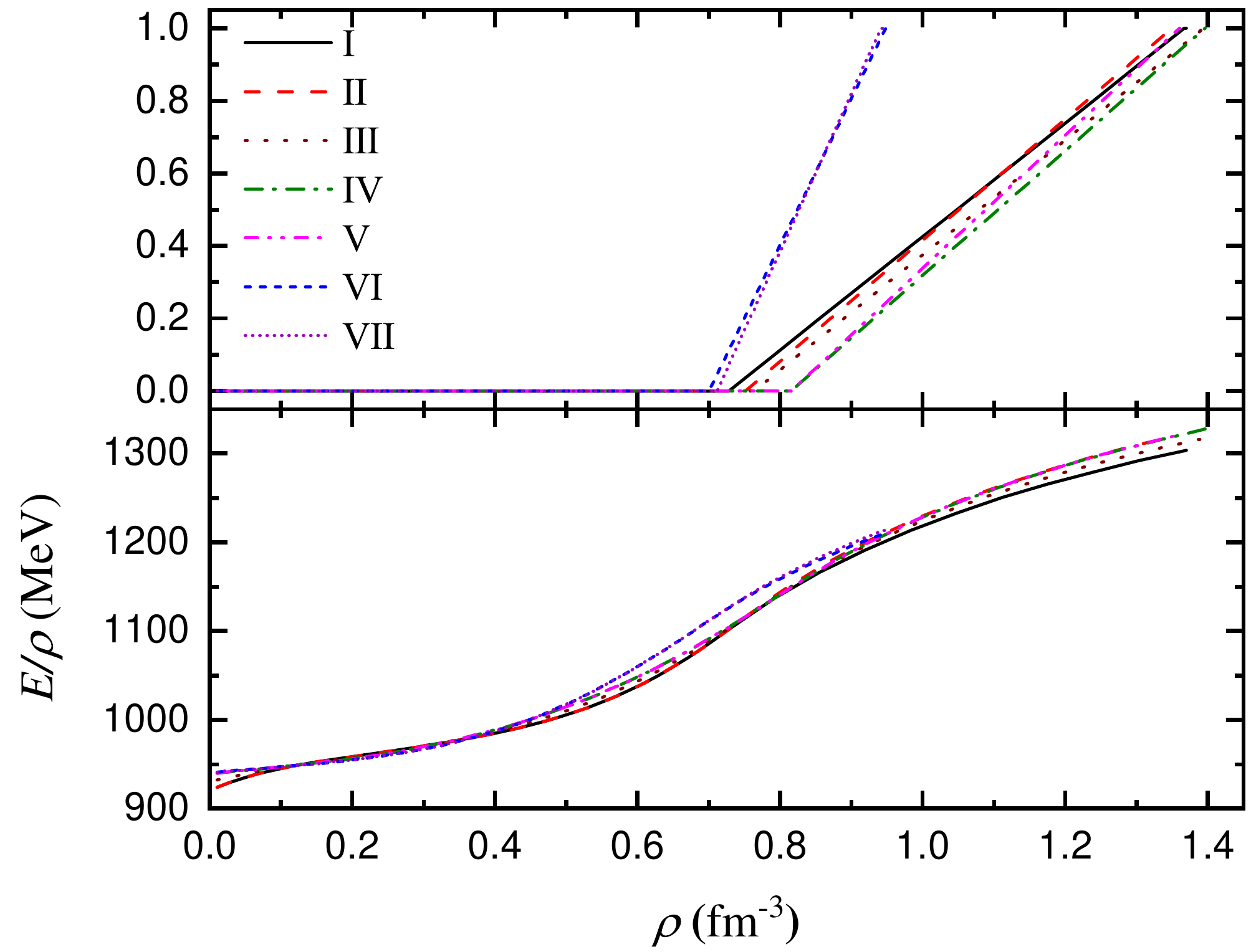}
  \caption{\label{Fig:EOS}The EOSs (left) and energy per baryon (right) of nuclear matter, quark matter, and their mixed phase inside hybrid stars that are consistent with pulsar observations, where the parameters sets indicated in Table~\ref{tableb} are adopted.}
\end{figure*}

For fixed baryon number densities, the energy density of nuclear matter, quark matter, and their mixed phase inside hybrid stars are obtained with Eqs.~(\ref{Eq:EH}), (\ref{Eq:EQ}), and (\ref{Eq:ET}), while the corresponding pressures are determined by Eqs.~(\ref{Eq:PH}), (\ref{Eq:PQ}), and (\ref{Eq:PT}). In the left panel of Fig.~\ref{Fig:EOS} we present the EOSs of stellar matter in hybrid stars that are consistent with pulsar observations, where the parameter sets indicated in Table~\ref{tableb} are adopted. We note that the EOSs for nuclear matter generally coincide with each other and are well constrained with the recent pulsar observations, while this is not the case for quark matter and quark-hadron mixed phase with much larger ambiguities. Since we have adopted the Maxwell construction, from the left panel of Fig.~\ref{Fig:EOS}, it is clearly seen that the pressure becomes constant for the quark-hadron mixed phase, where the relative energy density jump ranges from $\Delta E/E_\mathrm{t}\approx 1.4$ to 2.3. For the energy per baryon, as indicated in the right panel of Fig.~\ref{Fig:EOS}, it is increasing with density as the quark fraction increases from $\chi=0$ to 1. It is found that the onset density of deconfinement phase transition is relatively large, which ranges from $\rho_\mathrm{t}\approx$ 4.3 $\rho_0$ to 5.1 $\rho_0$. Meanwhile, the density of quark matter at the center of hybrid stars are even larger with $\rho^Q\approx$ 5.8-8.7 $\rho_0$.

\begin{figure*}[!ht]
  \centering
  \includegraphics[width=0.47\linewidth]{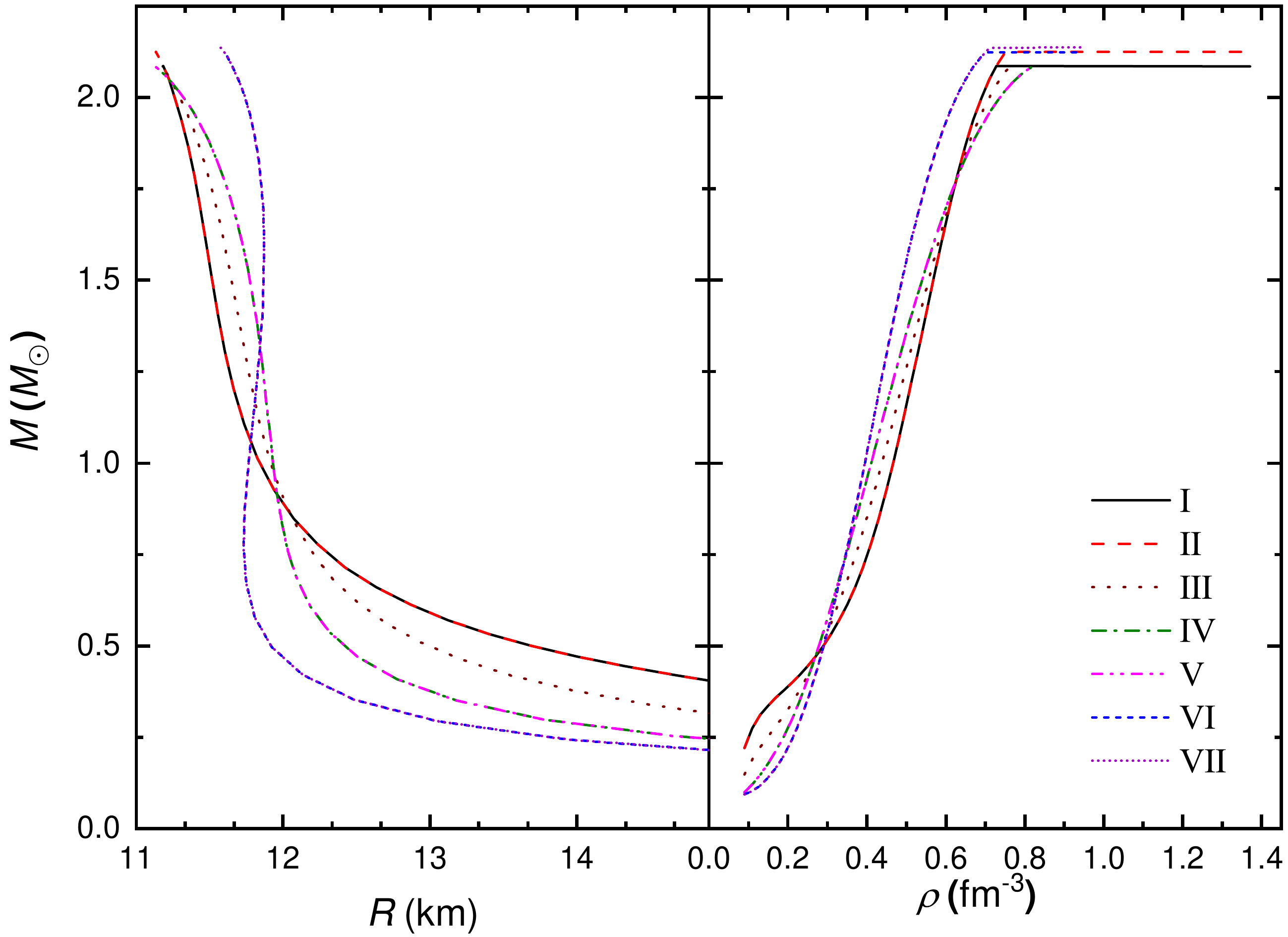}
  \caption{\label{Fig:MR} Mass-radius relations of hybrid stars obtained with the EOSs presented in Fig.~\ref{Fig:EOS}. The maximum masses $M_{max}$ and $R_{1.4}$ ($R_{2.08}$) of 1.4 $M_{\odot}$ (2.08 $M_{\odot}$) compact stars are indicated in Table~\ref{tableb}.}
\end{figure*}

Based on the EOSs indicated in Fig.~\ref{Fig:EOS}, the corresponding structures of hybrid stars are obtained by solving the Tolman-Oppenheimer-Volkov (TOV) equation
\begin{eqnarray}
&&\frac{\mbox{d}P}{\mbox{d}r} = -\frac{G M E}{r^2}   \frac{(1+P/E)(1+4\pi r^3 P/M)} {1-2G M/r},  \label{eq:TOV}\\
&&\frac{\mbox{d}M}{\mbox{d}r} = 4\pi E r^2, \label{eq:m_star}
\end{eqnarray}
where $G=6.707\times 10^{-45}\ \mathrm{MeV}^{-2}$ is the gravity constant. At the same time, the dimensionless tidal deformability is obtained with
\begin{equation}
\Lambda = \frac{2 k_2}{3}\left( \frac{R}{G M} \right)^5, \label{eq:td}
\end{equation}
where the second Love number $k_2$ is evaluated by introducing perturbations to the metric~\cite{Damour2009_PRD80-084035,Hinderer2010_PRD81-123016,Postnikov2010_PRD82-024016}. In Fig.~\ref{Fig:MR}, we present the masses of neutron stars as functions of radius (left panel) and central baryon number density (right panel), which are consistent with the recent astrophysical observations. The maximum masses and the radii of 1.4 $M_{\odot}$ (2.08 $M_{\odot}$) compact stars are indicated in Table~\ref{tableb}.  All the maximum masses of hybrid stars predicted by various EOSs in Fig.~\ref{Fig:EOS} are consistent with the observational mass. As quarks start to appear at the center of hybrid stars, the mass and radii become smaller and eventually hybrid stars become unstable.

\begin{figure}[!ht]
\centering
  \includegraphics[width=\linewidth]{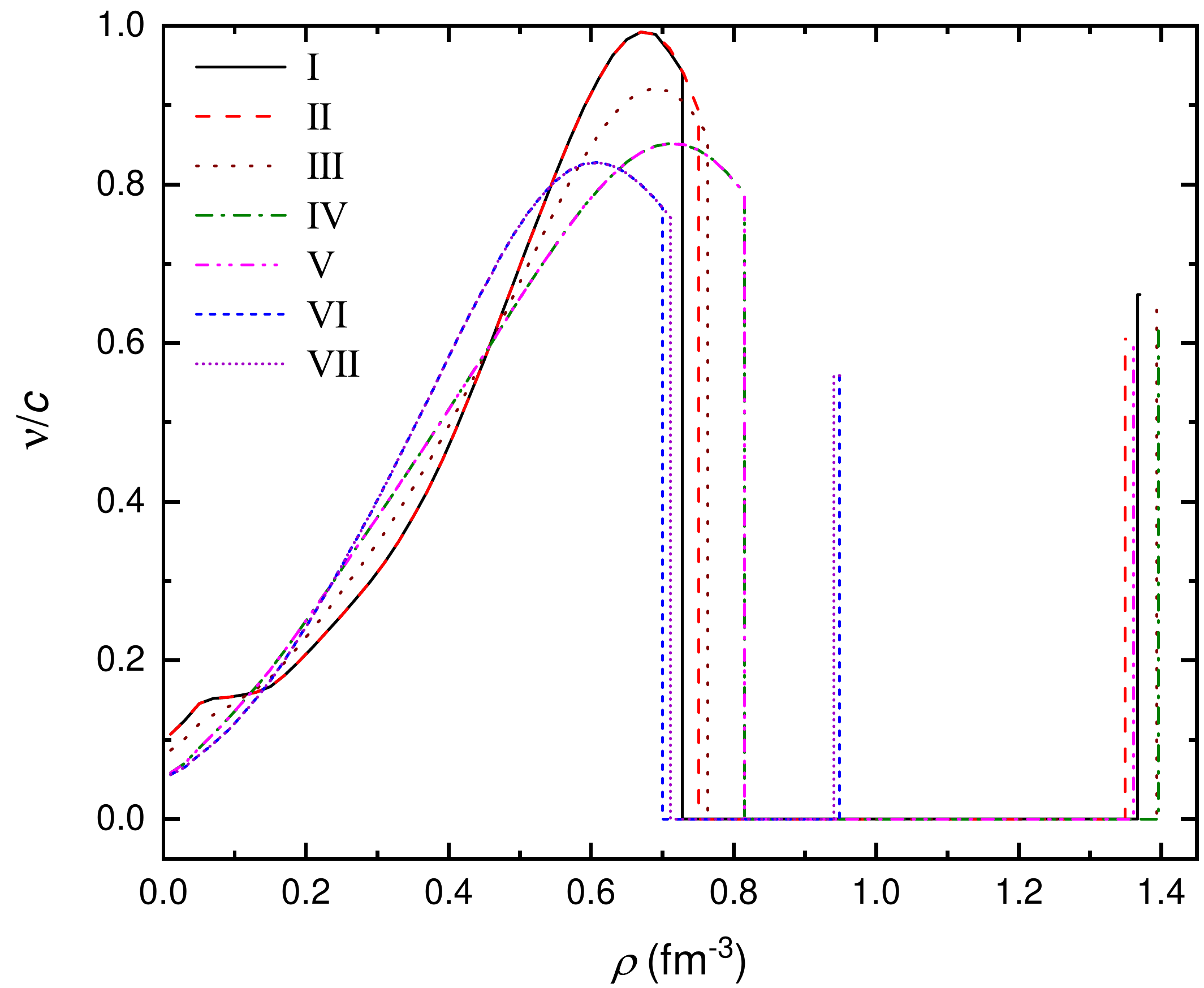}
  \caption{\label{Fig:v_all}The velocity of sound $v$ for hybrid star matter obtained with the EOSs presented in Fig.~\ref{Fig:EOS}.}
\end{figure}

Based on the results indicated in Fig.~\ref{Fig:EOS}, we can obtain the velocity of sound $v$ using the formula
\begin{equation}
  v = \sqrt{\frac{\mbox{d}P}{\mbox{d}E}}
\end{equation}
and present our results in Fig.~\ref{Fig:v_all}. As baryon number densities increases, the velocity of sound also increases at small densities. It is found that there exist a maximum for the velocity of sound at $\rho\approx 3.8$-$4.4 \rho_0$. In particular, we find $v$ increases until reaches its peak at $v=v_\mathrm{max}\approx 0.8 c$-$c$ and then decreases for nuclear matter. In the quark-hadron mixed phase, the velocity of sound $v$ vanishes, which suddenly increases once the pure quark phase takes place. Note that at exceedingly large densities, it is expected that $v$ approaches to the conformal limit $c/\sqrt{3}$ due to the asymptotic freedom of strong interaction~\cite{Annala2020_NP, Xia2014_PRD89-105027, Xia2021_CPC45-055104}.

\begin{figure}[!ht]
\centering
  \includegraphics[width=\linewidth]{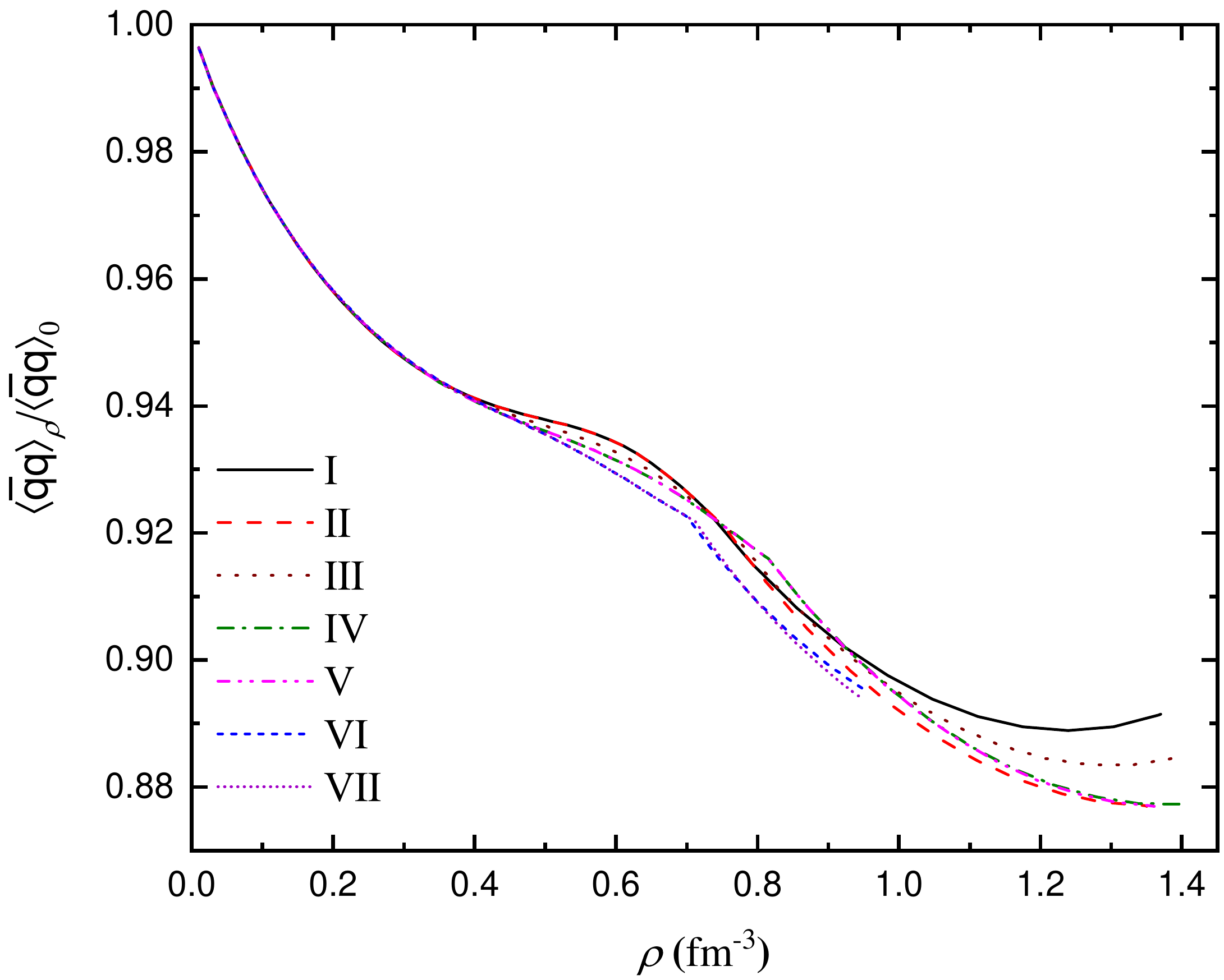}
  \caption{\label{Fig:cond_all}Relative quark condensate ${\langle\bar{q}q\rangle_{\rho}}/{\langle\bar{q}q\rangle_{0}}$ of hybrid star matter as functions of total baryon number density $\rho$, which correspond to the EOSs indicated in Fig.~\ref{Fig:EOS}.}
\end{figure}

Finally, based on the obtained energy density in Fig.~\ref{Fig:EOS}, the in-medium quark condensate for hybrid star matter can be fixed according to the equivparticle model elaborated in Sec.~\ref{sec:the_Equiv}. The obtained results are presented in Fig.~\ref{Fig:cond_all}. Since we have considered explicitly the first-order deconfinement phase transition in the framework of Maxwell construction, the in-medium quark condensate as a function of pressure $P$ or chemical potential $\mu_\mathrm{b}$ will decrease abruptly at the onset of the quark phase. In general, the relative quark condensate decreases nonlinearly with density and deviates from the expression~\cite{Cohen1992_PRC45-1881}
\begin{equation}
 \frac{\langle\bar{q}q\rangle_\rho} {\langle\bar{q}q\rangle_0}=1-\frac{\rho}{n^*}. \label{Eq:cond_lin}
\end{equation}
However, we find that for few cases the in-medium quark condensate increases at larger densities for the pure quark phase. Throughout the density range of hybrid stars ($\rho\leq\rho_\mathrm{TOV}$), the obtained quark condensate does not vanish. In such cases, even with an explicit deconfinement phase transition inside a hybrid star, the chiral symmetry of the stellar matter is only partially restored.

\section{\label{sec:con}Conclusion}
In this work, we investigate systematically the possible phase transition from nuclear matter to quark matter and its influence on hybrid star structures. For nuclear matter, the properties are obtained by carrying out a Taylor expansion of the binding energy to the order of $\rho^3$~\cite{Zhang2018_ApJ859-90}. For quark matter, we adopt an equivparticle model including both linear confinement and leading-order perturbative interactions. Based on the EOSs of nuclear matter and quark matter, their mixed phase and the corresponding EOSs for hybrid star matter are obtained using Maxwell construction. We then investigate the mass-radius relations of hybrid stars by solving the TOV equation, where the maximum mass and radii become smaller as quark matter starts to appear at the centers of hybrid stars. For those consistent with pulsar observations, it is found that the quark core is rather small and does not emerge for compact stars with $M \lesssim 2M_\odot$. We note that the relative energy density jump ranges from $\Delta E/E_\mathrm{t}\approx 1.4$ to 2.3 and  the onset density for deconfinement phase transition from $\rho_\mathrm{t}\approx$ 4.3 $\rho_0$ to 5.1 $\rho_0$. The velocity of sound $v$ reaches its peak at $\rho\approx 3.8$-$4.4\rho_0$. Meanwhile, even with the occurrence of a strong first-order deconfinement phase transition, we find that the velocity of sound still deviates from the conformal limit $c/\sqrt{3}$  at $\rho\approx\rho_\mathrm{TOV}$. Based on the constrained properties of dense stellar matter, we extract the corresponding quark condensate in the framework of equivparticle model~\cite{Peng2002_PLB548-189}, and find it is decreasing nonlinearly with density. At larger densities with pure quark matter, the quark condensate is still large and does not
decrease quickly with density, suggesting that the stellar matter in hybrid stars are highly nonperturbative even when a deconfinement phase transition takes place at $\rho \leqslant \rho_\mathrm{TOV}$, which is consistent with the recent studies in the absence of a strong first-order phase transition~\cite{Minamikawa2021,Jin2022}.

\begin{acknowledgments}
This work was partly supported by the National Natural Science Foundation of China (Grant Nos. U2032141, 11705163, 11875052), the National SKA Program of China (No. 2020SKA0120300), the Natural Science Foundation of Henan Province (202300410479), the Foundation of Fundamental Research for Young Teachers of Zhengzhou University (JC202041041), and the Physics Research and Development Program of Zhengzhou University (32410217).
\end{acknowledgments}


\begin{thebibliography}{67}%
\makeatletter
\providecommand \@ifxundefined [1]{%
 \@ifx{#1\undefined}
}%
\providecommand \@ifnum [1]{%
 \ifnum #1\expandafter \@firstoftwo
 \else \expandafter \@secondoftwo
 \fi
}%
\providecommand \@ifx [1]{%
 \ifx #1\expandafter \@firstoftwo
 \else \expandafter \@secondoftwo
 \fi
}%
\providecommand \natexlab [1]{#1}%
\providecommand \enquote  [1]{``#1''}%
\providecommand \bibnamefont  [1]{#1}%
\providecommand \bibfnamefont [1]{#1}%
\providecommand \citenamefont [1]{#1}%
\providecommand \href@noop [0]{\@secondoftwo}%
\providecommand \href [0]{\begingroup \@sanitize@url \@href}%
\providecommand \@href[1]{\@@startlink{#1}\@@href}%
\providecommand \@@href[1]{\endgroup#1\@@endlink}%
\providecommand \@sanitize@url [0]{\catcode `\\12\catcode `\$12\catcode
  `\&12\catcode `\#12\catcode `\^12\catcode `\_12\catcode `\%12\relax}%
\providecommand \@@startlink[1]{}%
\providecommand \@@endlink[0]{}%
\providecommand \url  [0]{\begingroup\@sanitize@url \@url }%
\providecommand \@url [1]{\endgroup\@href {#1}{\urlprefix }}%
\providecommand \urlprefix  [0]{URL }%
\providecommand \Eprint [0]{\href }%
\providecommand \doibase [0]{http://dx.doi.org/}%
\providecommand \selectlanguage [0]{\@gobble}%
\providecommand \bibinfo  [0]{\@secondoftwo}%
\providecommand \bibfield  [0]{\@secondoftwo}%
\providecommand \translation [1]{[#1]}%
\providecommand \BibitemOpen [0]{}%
\providecommand \bibitemStop [0]{}%
\providecommand \bibitemNoStop [0]{.\EOS\space}%
\providecommand \EOS [0]{\spacefactor3000\relax}%
\providecommand \BibitemShut  [1]{\csname bibitem#1\endcsname}%
\let\auto@bib@innerbib\@empty
\bibitem [{\citenamefont {{LIGO Scientific and Virgo
  Collaborations}}(2018)}]{LVC2018_PRL121-161101}%
  \BibitemOpen
  \bibfield  {author} {\bibinfo {author} {\bibnamefont {{LIGO Scientific and
  Virgo Collaborations}}},\ }\href {\doibase 10.1103/PhysRevLett.121.161101}
  {\bibfield  {journal} {\bibinfo  {journal} {Phys. Rev. Lett.}\ }\textbf
  {\bibinfo {volume} {121}},\ \bibinfo {pages} {161101} (\bibinfo {year}
  {2018})}\BibitemShut {NoStop}%
\bibitem [{\citenamefont {Bauswein}\ \emph {et~al.}(2019)\citenamefont
  {Bauswein}, \citenamefont {Bastian}, \citenamefont {Blaschke}, \citenamefont
  {Chatziioannou}, \citenamefont {Clark}, \citenamefont {Fischer},\ and\
  \citenamefont {Oertel}}]{Bauswein2019}%
  \BibitemOpen
  \bibfield  {author} {\bibinfo {author} {\bibfnamefont {A.}~\bibnamefont
  {Bauswein}}, \bibinfo {author} {\bibfnamefont {N.-U.~F.}\ \bibnamefont
  {Bastian}}, \bibinfo {author} {\bibfnamefont {D.~B.}\ \bibnamefont
  {Blaschke}}, \bibinfo {author} {\bibfnamefont {K.}~\bibnamefont
  {Chatziioannou}}, \bibinfo {author} {\bibfnamefont {J.~A.}\ \bibnamefont
  {Clark}}, \bibinfo {author} {\bibfnamefont {T.}~\bibnamefont {Fischer}}, \
  and\ \bibinfo {author} {\bibfnamefont {M.}~\bibnamefont {Oertel}},\ }\href
  {\doibase https://doi.org/10.1103/PhysRevLett.122.061102} {\bibfield
  {journal} {\bibinfo  {journal} {Phys. Rev. Lett}\ }\textbf {\bibinfo {volume}
  {122}},\ \bibinfo {pages} {061102} (\bibinfo {year} {2019})}\BibitemShut
  {NoStop}%
\bibitem [{\citenamefont {Most}\ \emph {et~al.}(2019)\citenamefont {Most},
  \citenamefont {Papenfort}, \citenamefont {Dexheimer}, \citenamefont
  {Hanauske}, \citenamefont {Schramm}, \citenamefont {St{\"o}cker},\ and\
  \citenamefont {Rezzolla}}]{Most2019}%
  \BibitemOpen
  \bibfield  {author} {\bibinfo {author} {\bibfnamefont {E.~R.}\ \bibnamefont
  {Most}}, \bibinfo {author} {\bibfnamefont {L.~J.}\ \bibnamefont {Papenfort}},
  \bibinfo {author} {\bibfnamefont {V.}~\bibnamefont {Dexheimer}}, \bibinfo
  {author} {\bibfnamefont {M.}~\bibnamefont {Hanauske}}, \bibinfo {author}
  {\bibfnamefont {S.}~\bibnamefont {Schramm}}, \bibinfo {author} {\bibfnamefont
  {H.}~\bibnamefont {St{\"o}cker}}, \ and\ \bibinfo {author} {\bibfnamefont
  {L.}~\bibnamefont {Rezzolla}},\ }\href {\doibase
  10.1103/physrevlett.122.061101} {\bibfield  {journal} {\bibinfo  {journal}
  {Phys. Rev. Lett}\ }\textbf {\bibinfo {volume} {122}},\ \bibinfo {pages}
  {061101} (\bibinfo {year} {2019})}\BibitemShut {NoStop}%
\bibitem [{\citenamefont {Riley}\ \emph {et~al.}(2019)\citenamefont {Riley},
  \citenamefont {Watts}, \citenamefont {Bogdanov}, \citenamefont {Ray},
  \citenamefont {Ludlam}, \citenamefont {Guillot}, \citenamefont {Arzoumanian},
  \citenamefont {Baker}, \citenamefont {Bilous}, \citenamefont {Chakrabarty},
  \citenamefont {Gendreau}, \citenamefont {Harding}, \citenamefont {Ho},
  \citenamefont {Lattimer}, \citenamefont {Morsink},\ and\ \citenamefont
  {Strohmayer}}]{Riley2019_ApJ887-L21}%
  \BibitemOpen
  \bibfield  {author} {\bibinfo {author} {\bibfnamefont {T.~E.}\ \bibnamefont
  {Riley}}, \bibinfo {author} {\bibfnamefont {A.~L.}\ \bibnamefont {Watts}},
  \bibinfo {author} {\bibfnamefont {S.}~\bibnamefont {Bogdanov}}, \bibinfo
  {author} {\bibfnamefont {P.~S.}\ \bibnamefont {Ray}}, \bibinfo {author}
  {\bibfnamefont {R.~M.}\ \bibnamefont {Ludlam}}, \bibinfo {author}
  {\bibfnamefont {S.}~\bibnamefont {Guillot}}, \bibinfo {author} {\bibfnamefont
  {Z.}~\bibnamefont {Arzoumanian}}, \bibinfo {author} {\bibfnamefont {C.~L.}\
  \bibnamefont {Baker}}, \bibinfo {author} {\bibfnamefont {A.~V.}\ \bibnamefont
  {Bilous}}, \bibinfo {author} {\bibfnamefont {D.}~\bibnamefont {Chakrabarty}},
  \bibinfo {author} {\bibfnamefont {K.~C.}\ \bibnamefont {Gendreau}}, \bibinfo
  {author} {\bibfnamefont {A.~K.}\ \bibnamefont {Harding}}, \bibinfo {author}
  {\bibfnamefont {W.~C.~G.}\ \bibnamefont {Ho}}, \bibinfo {author}
  {\bibfnamefont {J.~M.}\ \bibnamefont {Lattimer}}, \bibinfo {author}
  {\bibfnamefont {S.~M.}\ \bibnamefont {Morsink}}, \ and\ \bibinfo {author}
  {\bibfnamefont {T.~E.}\ \bibnamefont {Strohmayer}},\ }\href {\doibase
  10.3847/2041-8213/ab481c} {\bibfield  {journal} {\bibinfo  {journal}
  {Astrophys. J.}\ }\textbf {\bibinfo {volume} {887}},\ \bibinfo {pages} {L21}
  (\bibinfo {year} {2019})}\BibitemShut {NoStop}%
\bibitem [{\citenamefont {Riley}\ \emph {et~al.}(2021)\citenamefont {Riley},
  \citenamefont {Watts}, \citenamefont {Ray}, \citenamefont {Bogdanov},
  \citenamefont {Guillot}, \citenamefont {Morsink}, \citenamefont {Bilous},
  \citenamefont {Arzoumanian}, \citenamefont {Choudhury}, \citenamefont
  {Deneva} \emph {et~al.}}]{Riley2021}%
  \BibitemOpen
  \bibfield  {author} {\bibinfo {author} {\bibfnamefont {T.~E.}\ \bibnamefont
  {Riley}}, \bibinfo {author} {\bibfnamefont {A.~L.}\ \bibnamefont {Watts}},
  \bibinfo {author} {\bibfnamefont {P.~S.}\ \bibnamefont {Ray}}, \bibinfo
  {author} {\bibfnamefont {S.}~\bibnamefont {Bogdanov}}, \bibinfo {author}
  {\bibfnamefont {S.}~\bibnamefont {Guillot}}, \bibinfo {author} {\bibfnamefont
  {S.~M.}\ \bibnamefont {Morsink}}, \bibinfo {author} {\bibfnamefont {A.~V.}\
  \bibnamefont {Bilous}}, \bibinfo {author} {\bibfnamefont {Z.}~\bibnamefont
  {Arzoumanian}}, \bibinfo {author} {\bibfnamefont {D.}~\bibnamefont
  {Choudhury}}, \bibinfo {author} {\bibfnamefont {J.~S.}\ \bibnamefont
  {Deneva}},  \emph {et~al.},\ }\href {\doibase
  https://doi.org/10.3847/2041-8213/ac0a81} {\bibfield  {journal} {\bibinfo
  {journal} {Astrophys. J.}\ }\textbf {\bibinfo {volume} {918}},\ \bibinfo
  {pages} {L27} (\bibinfo {year} {2021})}\BibitemShut {NoStop}%
\bibitem [{\citenamefont {Miller}\ \emph {et~al.}(2019)\citenamefont {Miller},
  \citenamefont {Lamb}, \citenamefont {Dittmann}, \citenamefont {Bogdanov},
  \citenamefont {Arzoumanian}, \citenamefont {Gendreau}, \citenamefont
  {Guillot}, \citenamefont {Harding}, \citenamefont {Ho}, \citenamefont
  {Lattimer}, \citenamefont {Ludlam}, \citenamefont {Mahmoodifar},
  \citenamefont {Morsink}, \citenamefont {Ray}, \citenamefont {Strohmayer},
  \citenamefont {Wood}, \citenamefont {Enoto}, \citenamefont {Foster},
  \citenamefont {Okajima}, \citenamefont {Prigozhin},\ and\ \citenamefont
  {Soong}}]{Miller2019_ApJ887-L24}%
  \BibitemOpen
  \bibfield  {author} {\bibinfo {author} {\bibfnamefont {M.~C.}\ \bibnamefont
  {Miller}}, \bibinfo {author} {\bibfnamefont {F.~K.}\ \bibnamefont {Lamb}},
  \bibinfo {author} {\bibfnamefont {A.~J.}\ \bibnamefont {Dittmann}}, \bibinfo
  {author} {\bibfnamefont {S.}~\bibnamefont {Bogdanov}}, \bibinfo {author}
  {\bibfnamefont {Z.}~\bibnamefont {Arzoumanian}}, \bibinfo {author}
  {\bibfnamefont {K.~C.}\ \bibnamefont {Gendreau}}, \bibinfo {author}
  {\bibfnamefont {S.}~\bibnamefont {Guillot}}, \bibinfo {author} {\bibfnamefont
  {A.~K.}\ \bibnamefont {Harding}}, \bibinfo {author} {\bibfnamefont
  {W.~C.~G.}\ \bibnamefont {Ho}}, \bibinfo {author} {\bibfnamefont {J.~M.}\
  \bibnamefont {Lattimer}}, \bibinfo {author} {\bibfnamefont {R.~M.}\
  \bibnamefont {Ludlam}}, \bibinfo {author} {\bibfnamefont {S.}~\bibnamefont
  {Mahmoodifar}}, \bibinfo {author} {\bibfnamefont {S.~M.}\ \bibnamefont
  {Morsink}}, \bibinfo {author} {\bibfnamefont {P.~S.}\ \bibnamefont {Ray}},
  \bibinfo {author} {\bibfnamefont {T.~E.}\ \bibnamefont {Strohmayer}},
  \bibinfo {author} {\bibfnamefont {K.~S.}\ \bibnamefont {Wood}}, \bibinfo
  {author} {\bibfnamefont {T.}~\bibnamefont {Enoto}}, \bibinfo {author}
  {\bibfnamefont {R.}~\bibnamefont {Foster}}, \bibinfo {author} {\bibfnamefont
  {T.}~\bibnamefont {Okajima}}, \bibinfo {author} {\bibfnamefont
  {G.}~\bibnamefont {Prigozhin}}, \ and\ \bibinfo {author} {\bibfnamefont
  {Y.}~\bibnamefont {Soong}},\ }\href {\doibase 10.3847/2041-8213/ab50c5}
  {\bibfield  {journal} {\bibinfo  {journal} {Astrophys. J.}\ }\textbf
  {\bibinfo {volume} {887}},\ \bibinfo {pages} {L24} (\bibinfo {year}
  {2019})}\BibitemShut {NoStop}%
\bibitem [{\citenamefont {Miller}\ \emph {et~al.}(2021)\citenamefont {Miller},
  \citenamefont {Lamb}, \citenamefont {Dittmann}, \citenamefont {Bogdanov},
  \citenamefont {Arzoumanian}, \citenamefont {Gendreau}, \citenamefont
  {Guillot}, \citenamefont {Ho}, \citenamefont {Lattimer}, \citenamefont
  {Loewenstein} \emph {et~al.}}]{Miller2021}%
  \BibitemOpen
  \bibfield  {author} {\bibinfo {author} {\bibfnamefont {M.}~\bibnamefont
  {Miller}}, \bibinfo {author} {\bibfnamefont {F.}~\bibnamefont {Lamb}},
  \bibinfo {author} {\bibfnamefont {A.}~\bibnamefont {Dittmann}}, \bibinfo
  {author} {\bibfnamefont {S.}~\bibnamefont {Bogdanov}}, \bibinfo {author}
  {\bibfnamefont {Z.}~\bibnamefont {Arzoumanian}}, \bibinfo {author}
  {\bibfnamefont {K.}~\bibnamefont {Gendreau}}, \bibinfo {author}
  {\bibfnamefont {S.}~\bibnamefont {Guillot}}, \bibinfo {author} {\bibfnamefont
  {W.}~\bibnamefont {Ho}}, \bibinfo {author} {\bibfnamefont {J.}~\bibnamefont
  {Lattimer}}, \bibinfo {author} {\bibfnamefont {M.}~\bibnamefont
  {Loewenstein}},  \emph {et~al.},\ }\href
  {https://doi.org/10.3847/2041-8213/ac089b} {\bibfield  {journal} {\bibinfo
  {journal} {Astrophys. J.}\ }\textbf {\bibinfo {volume} {918}},\ \bibinfo
  {pages} {L28} (\bibinfo {year} {2021})}\BibitemShut {NoStop}%
\bibitem [{\citenamefont {Li}\ \emph {et~al.}(2019)\citenamefont {Li},
  \citenamefont {Krastev}, \citenamefont {Wen},\ and\ \citenamefont
  {Zhang}}]{Li2019}%
  \BibitemOpen
  \bibfield  {author} {\bibinfo {author} {\bibfnamefont {B.-A.}\ \bibnamefont
  {Li}}, \bibinfo {author} {\bibfnamefont {P.~G.}\ \bibnamefont {Krastev}},
  \bibinfo {author} {\bibfnamefont {D.-H.}\ \bibnamefont {Wen}}, \ and\
  \bibinfo {author} {\bibfnamefont {N.-B.}\ \bibnamefont {Zhang}},\ }\href
  {\doibase https://doi.org/10.1140/epja/i2019-12780-8} {\bibfield  {journal}
  {\bibinfo  {journal} {Eur. Phys. J. A}\ }\textbf {\bibinfo {volume} {55}},\
  \bibinfo {pages} {1} (\bibinfo {year} {2019})}\BibitemShut {NoStop}%
\bibitem [{\citenamefont {Annala}\ \emph
  {et~al.}(2020{\natexlab{a}})\citenamefont {Annala}, \citenamefont {Gorda},
  \citenamefont {Kurkela}, \citenamefont {N{\"a}ttil{\"a}},\ and\ \citenamefont
  {Vuorinen}}]{Annala2020}%
  \BibitemOpen
  \bibfield  {author} {\bibinfo {author} {\bibfnamefont {E.}~\bibnamefont
  {Annala}}, \bibinfo {author} {\bibfnamefont {T.}~\bibnamefont {Gorda}},
  \bibinfo {author} {\bibfnamefont {A.}~\bibnamefont {Kurkela}}, \bibinfo
  {author} {\bibfnamefont {J.}~\bibnamefont {N{\"a}ttil{\"a}}}, \ and\ \bibinfo
  {author} {\bibfnamefont {A.}~\bibnamefont {Vuorinen}},\ }\href {\doibase
  https://doi.org/10.1038/s41567-020-0914-9} {\bibfield  {journal} {\bibinfo
  {journal} {Nature Physics}\ }\textbf {\bibinfo {volume} {16}},\ \bibinfo
  {pages} {907} (\bibinfo {year} {2020}{\natexlab{a}})}\BibitemShut {NoStop}%
\bibitem [{\citenamefont {Alford}\ \emph {et~al.}(2007)\citenamefont {Alford},
  \citenamefont {Blaschke}, \citenamefont {Drago}, \citenamefont {Kl{\"a}hn},
  \citenamefont {Pagliara},\ and\ \citenamefont
  {Schaffner-Bielich}}]{Alford2007}%
  \BibitemOpen
  \bibfield  {author} {\bibinfo {author} {\bibfnamefont {M.}~\bibnamefont
  {Alford}}, \bibinfo {author} {\bibfnamefont {D.}~\bibnamefont {Blaschke}},
  \bibinfo {author} {\bibfnamefont {A.}~\bibnamefont {Drago}}, \bibinfo
  {author} {\bibfnamefont {T.}~\bibnamefont {Kl{\"a}hn}}, \bibinfo {author}
  {\bibfnamefont {G.}~\bibnamefont {Pagliara}}, \ and\ \bibinfo {author}
  {\bibfnamefont {J.}~\bibnamefont {Schaffner-Bielich}},\ }\href {\doibase
  https://doi.org/10.1038/nature05582} {\bibfield  {journal} {\bibinfo
  {journal} {Nature}\ }\textbf {\bibinfo {volume} {445}},\ \bibinfo {pages}
  {E7} (\bibinfo {year} {2007})}\BibitemShut {NoStop}%
\bibitem [{\citenamefont {Weber}(2017)}]{Weber2017}%
  \BibitemOpen
  \bibfield  {author} {\bibinfo {author} {\bibfnamefont {F.}~\bibnamefont
  {Weber}},\ }\href {\doibase https://doi.org/10.1201/9780203741719} {\emph
  {\bibinfo {title} {Pulsars as astrophysical laboratories for nuclear and
  particle physics}}}\ (\bibinfo  {publisher} {Routledge},\ \bibinfo {year}
  {2017})\BibitemShut {NoStop}%
\bibitem [{\citenamefont {Maruyama}\ \emph {et~al.}(1994)\citenamefont
  {Maruyama}, \citenamefont {Fujii}, \citenamefont {Muto},\ and\ \citenamefont
  {Tatsumi}}]{Maruyama1994_PLB337-19}%
  \BibitemOpen
  \bibfield  {author} {\bibinfo {author} {\bibfnamefont {T.}~\bibnamefont
  {Maruyama}}, \bibinfo {author} {\bibfnamefont {H.}~\bibnamefont {Fujii}},
  \bibinfo {author} {\bibfnamefont {T.}~\bibnamefont {Muto}}, \ and\ \bibinfo
  {author} {\bibfnamefont {T.}~\bibnamefont {Tatsumi}},\ }\href {\doibase
  https://doi.org/10.1016/0370-2693(94)91436-2} {\bibfield  {journal} {\bibinfo
   {journal} {Phys. Lett. B}\ }\textbf {\bibinfo {volume} {337}},\ \bibinfo
  {pages} {19 } (\bibinfo {year} {1994})}\BibitemShut {NoStop}%
\bibitem [{\citenamefont {Maruyama}\ \emph {et~al.}(2006)\citenamefont
  {Maruyama}, \citenamefont {Tatsumi}, \citenamefont {Voskresensky},
  \citenamefont {Tanigawa}, \citenamefont {Endo},\ and\ \citenamefont
  {Chiba}}]{Maruyama2006_PRC73-035802}%
  \BibitemOpen
  \bibfield  {author} {\bibinfo {author} {\bibfnamefont {T.}~\bibnamefont
  {Maruyama}}, \bibinfo {author} {\bibfnamefont {T.}~\bibnamefont {Tatsumi}},
  \bibinfo {author} {\bibfnamefont {D.~N.}\ \bibnamefont {Voskresensky}},
  \bibinfo {author} {\bibfnamefont {T.}~\bibnamefont {Tanigawa}}, \bibinfo
  {author} {\bibfnamefont {T.}~\bibnamefont {Endo}}, \ and\ \bibinfo {author}
  {\bibfnamefont {S.}~\bibnamefont {Chiba}},\ }\href {\doibase
  10.1103/PhysRevC.73.035802} {\bibfield  {journal} {\bibinfo  {journal} {Phys.
  Rev. C}\ }\textbf {\bibinfo {volume} {73}},\ \bibinfo {pages} {035802}
  (\bibinfo {year} {2006})}\BibitemShut {NoStop}%
\bibitem [{\citenamefont {Glendenning}(1992)}]{Glendenning1992}%
  \BibitemOpen
  \bibfield  {author} {\bibinfo {author} {\bibfnamefont {N.~K.}\ \bibnamefont
  {Glendenning}},\ }\href {\doibase https://doi.org/10.1103/PhysRevD.46.1274}
  {\bibfield  {journal} {\bibinfo  {journal} {Phys. Rev. D}\ }\textbf {\bibinfo
  {volume} {46}},\ \bibinfo {pages} {1274} (\bibinfo {year}
  {1992})}\BibitemShut {NoStop}%
\bibitem [{\citenamefont {Peng}\ \emph {et~al.}(2008)\citenamefont {Peng},
  \citenamefont {Li},\ and\ \citenamefont {Lombardo}}]{Peng2008_PRC77-065807}%
  \BibitemOpen
  \bibfield  {author} {\bibinfo {author} {\bibfnamefont {G.~X.}\ \bibnamefont
  {Peng}}, \bibinfo {author} {\bibfnamefont {A.}~\bibnamefont {Li}}, \ and\
  \bibinfo {author} {\bibfnamefont {U.}~\bibnamefont {Lombardo}},\ }\href
  {\doibase 10.1103/PhysRevC.77.065807} {\bibfield  {journal} {\bibinfo
  {journal} {Phys. Rev. C}\ }\textbf {\bibinfo {volume} {77}},\ \bibinfo
  {pages} {065807} (\bibinfo {year} {2008})}\BibitemShut {NoStop}%
\bibitem [{\citenamefont {Li}\ \emph {et~al.}(2015)\citenamefont {Li},
  \citenamefont {Zuo},\ and\ \citenamefont {Peng}}]{Li2015_PRC91-035803}%
  \BibitemOpen
  \bibfield  {author} {\bibinfo {author} {\bibfnamefont {A.}~\bibnamefont
  {Li}}, \bibinfo {author} {\bibfnamefont {W.}~\bibnamefont {Zuo}}, \ and\
  \bibinfo {author} {\bibfnamefont {G.~X.}\ \bibnamefont {Peng}},\ }\href
  {\doibase 10.1103/PhysRevC.91.035803} {\bibfield  {journal} {\bibinfo
  {journal} {Phys. Rev. C}\ }\textbf {\bibinfo {volume} {91}},\ \bibinfo
  {pages} {035803} (\bibinfo {year} {2015})}\BibitemShut {NoStop}%
\bibitem [{\citenamefont {Kl\"ahn}\ \emph {et~al.}(2013)\citenamefont
  {Kl\"ahn}, \citenamefont {\L{}astowiecki},\ and\ \citenamefont
  {Blaschke}}]{Klahn2013_PRD88-085001}%
  \BibitemOpen
  \bibfield  {author} {\bibinfo {author} {\bibfnamefont {T.}~\bibnamefont
  {Kl\"ahn}}, \bibinfo {author} {\bibfnamefont {R.}~\bibnamefont
  {\L{}astowiecki}}, \ and\ \bibinfo {author} {\bibfnamefont {D.}~\bibnamefont
  {Blaschke}},\ }\href {\doibase 10.1103/PhysRevD.88.085001} {\bibfield
  {journal} {\bibinfo  {journal} {Phys. Rev. D}\ }\textbf {\bibinfo {volume}
  {88}},\ \bibinfo {pages} {085001} (\bibinfo {year} {2013})}\BibitemShut
  {NoStop}%
\bibitem [{\citenamefont {Bombaci}\ and\ \citenamefont
  {Logoteta}(2017)}]{Bombaci2016_IJMPD-1730004}%
  \BibitemOpen
  \bibfield  {author} {\bibinfo {author} {\bibfnamefont {I.}~\bibnamefont
  {Bombaci}}\ and\ \bibinfo {author} {\bibfnamefont {D.}~\bibnamefont
  {Logoteta}},\ }\href {\doibase 10.1142/S021827181730004X} {\bibfield
  {journal} {\bibinfo  {journal} {Int. J. Mod. Phys. D}\ ,\ \bibinfo {pages}
  {1730004}} (\bibinfo {year} {2017})}\BibitemShut {NoStop}%
\bibitem [{\citenamefont {Glendenning}\ and\ \citenamefont
  {Schaffner-Bielich}(1998)}]{Glendenning1998}%
  \BibitemOpen
  \bibfield  {author} {\bibinfo {author} {\bibfnamefont {N.~K.}\ \bibnamefont
  {Glendenning}}\ and\ \bibinfo {author} {\bibfnamefont {J.}~\bibnamefont
  {Schaffner-Bielich}},\ }\href {\doibase
  https://doi.org/10.1103/PhysRevLett.81.4564} {\bibfield  {journal} {\bibinfo
  {journal} {Phys. Rev. Lett}\ }\textbf {\bibinfo {volume} {81}},\ \bibinfo
  {pages} {4564} (\bibinfo {year} {1998})}\BibitemShut {NoStop}%
\bibitem [{\citenamefont {Glendenning}\ and\ \citenamefont
  {Schaffner-Bielich}(1999)}]{Glendenning1999}%
  \BibitemOpen
  \bibfield  {author} {\bibinfo {author} {\bibfnamefont {N.~K.}\ \bibnamefont
  {Glendenning}}\ and\ \bibinfo {author} {\bibfnamefont {J.}~\bibnamefont
  {Schaffner-Bielich}},\ }\href {\doibase
  https://doi.org/10.1103/PhysRevC.60.025803} {\bibfield  {journal} {\bibinfo
  {journal} {Phys. Rev. C}\ }\textbf {\bibinfo {volume} {60}},\ \bibinfo
  {pages} {025803} (\bibinfo {year} {1999})}\BibitemShut {NoStop}%
\bibitem [{\citenamefont {Heiselberg}\ \emph {et~al.}(1993)\citenamefont
  {Heiselberg}, \citenamefont {Pethick},\ and\ \citenamefont
  {Staubo}}]{Heiselberg1993_PRL70-1355}%
  \BibitemOpen
  \bibfield  {author} {\bibinfo {author} {\bibfnamefont {H.}~\bibnamefont
  {Heiselberg}}, \bibinfo {author} {\bibfnamefont {C.~J.}\ \bibnamefont
  {Pethick}}, \ and\ \bibinfo {author} {\bibfnamefont {E.~F.}\ \bibnamefont
  {Staubo}},\ }\href {\doibase 10.1103/PhysRevLett.70.1355} {\bibfield
  {journal} {\bibinfo  {journal} {Phys. Rev. Lett.}\ }\textbf {\bibinfo
  {volume} {70}},\ \bibinfo {pages} {1355} (\bibinfo {year}
  {1993})}\BibitemShut {NoStop}%
\bibitem [{\citenamefont {Voskresensky}\ \emph {et~al.}(2002)\citenamefont
  {Voskresensky}, \citenamefont {Yasuhira},\ and\ \citenamefont
  {Tatsumi}}]{Voskresensky2002_PLB541-93}%
  \BibitemOpen
  \bibfield  {author} {\bibinfo {author} {\bibfnamefont {D.}~\bibnamefont
  {Voskresensky}}, \bibinfo {author} {\bibfnamefont {M.}~\bibnamefont
  {Yasuhira}}, \ and\ \bibinfo {author} {\bibfnamefont {T.}~\bibnamefont
  {Tatsumi}},\ }\href
  {http://www.sciencedirect.com/science/article/pii/S037026930202186X}
  {\bibfield  {journal} {\bibinfo  {journal} {Phys. Lett. B}\ }\textbf
  {\bibinfo {volume} {541}},\ \bibinfo {pages} {93 } (\bibinfo {year}
  {2002})}\BibitemShut {NoStop}%
\bibitem [{\citenamefont {Tatsumi}\ \emph {et~al.}(2003)\citenamefont
  {Tatsumi}, \citenamefont {Yasuhira},\ and\ \citenamefont
  {Voskresensky}}]{Tatsumi2003_NPA718-359}%
  \BibitemOpen
  \bibfield  {author} {\bibinfo {author} {\bibfnamefont {T.}~\bibnamefont
  {Tatsumi}}, \bibinfo {author} {\bibfnamefont {M.}~\bibnamefont {Yasuhira}}, \
  and\ \bibinfo {author} {\bibfnamefont {D.}~\bibnamefont {Voskresensky}},\
  }\href {http://www.sciencedirect.com/science/article/pii/S0375947403007395}
  {\bibfield  {journal} {\bibinfo  {journal} {Nucl. Phys. A}\ }\textbf
  {\bibinfo {volume} {718}},\ \bibinfo {pages} {359 } (\bibinfo {year}
  {2003})}\BibitemShut {NoStop}%
\bibitem [{\citenamefont {Voskresensky}\ \emph {et~al.}(2003)\citenamefont
  {Voskresensky}, \citenamefont {Yasuhira},\ and\ \citenamefont
  {Tatsumi}}]{Voskresensky2003_NPA723-291}%
  \BibitemOpen
  \bibfield  {author} {\bibinfo {author} {\bibfnamefont {D.}~\bibnamefont
  {Voskresensky}}, \bibinfo {author} {\bibfnamefont {M.}~\bibnamefont
  {Yasuhira}}, \ and\ \bibinfo {author} {\bibfnamefont {T.}~\bibnamefont
  {Tatsumi}},\ }\href
  {http://www.sciencedirect.com/science/article/pii/S0375947403013137}
  {\bibfield  {journal} {\bibinfo  {journal} {Nucl. Phys. A}\ }\textbf
  {\bibinfo {volume} {723}},\ \bibinfo {pages} {291 } (\bibinfo {year}
  {2003})}\BibitemShut {NoStop}%
\bibitem [{\citenamefont {Bejger}\ \emph {et~al.}(2005)\citenamefont {Bejger},
  \citenamefont {Haensel},\ and\ \citenamefont {Zdunik}}]{Bejger2005}%
  \BibitemOpen
  \bibfield  {author} {\bibinfo {author} {\bibfnamefont {M.}~\bibnamefont
  {Bejger}}, \bibinfo {author} {\bibfnamefont {P.}~\bibnamefont {Haensel}}, \
  and\ \bibinfo {author} {\bibfnamefont {J.}~\bibnamefont {Zdunik}},\ }\href
  {https://doi.org/10.1111/j.1365-2966.2005.08933.x} {\bibfield  {journal}
  {\bibinfo  {journal} {Mon. Not. R. Astron. Soc}\ }\textbf {\bibinfo {volume}
  {359}},\ \bibinfo {pages} {699} (\bibinfo {year} {2005})}\BibitemShut
  {NoStop}%
\bibitem [{\citenamefont {Endo}\ \emph {et~al.}(2005)\citenamefont {Endo},
  \citenamefont {Maruyama}, \citenamefont {Chiba},\ and\ \citenamefont
  {Tatsumi}}]{Endo2005_NPA749-333}%
  \BibitemOpen
  \bibfield  {author} {\bibinfo {author} {\bibfnamefont {T.}~\bibnamefont
  {Endo}}, \bibinfo {author} {\bibfnamefont {T.}~\bibnamefont {Maruyama}},
  \bibinfo {author} {\bibfnamefont {S.}~\bibnamefont {Chiba}}, \ and\ \bibinfo
  {author} {\bibfnamefont {T.}~\bibnamefont {Tatsumi}},\ }\href
  {http://www.sciencedirect.com/science/article/pii/S0375947404012813}
  {\bibfield  {journal} {\bibinfo  {journal} {Nucl. Phys. A}\ }\textbf
  {\bibinfo {volume} {749}},\ \bibinfo {pages} {333} (\bibinfo {year}
  {2005})}\BibitemShut {NoStop}%
\bibitem [{\citenamefont {Maruyama}\ \emph {et~al.}(2007)\citenamefont
  {Maruyama}, \citenamefont {Chiba}, \citenamefont {Schulze},\ and\
  \citenamefont {Tatsumi}}]{Maruyama2007_PRD76-123015}%
  \BibitemOpen
  \bibfield  {author} {\bibinfo {author} {\bibfnamefont {T.}~\bibnamefont
  {Maruyama}}, \bibinfo {author} {\bibfnamefont {S.}~\bibnamefont {Chiba}},
  \bibinfo {author} {\bibfnamefont {H.-J.}\ \bibnamefont {Schulze}}, \ and\
  \bibinfo {author} {\bibfnamefont {T.}~\bibnamefont {Tatsumi}},\ }\href
  {\doibase 10.1103/PhysRevD.76.123015} {\bibfield  {journal} {\bibinfo
  {journal} {Phys. Rev. D}\ }\textbf {\bibinfo {volume} {76}},\ \bibinfo
  {pages} {123015} (\bibinfo {year} {2007})}\BibitemShut {NoStop}%
\bibitem [{\citenamefont {Yasutake}\ \emph {et~al.}(2014)\citenamefont
  {Yasutake}, \citenamefont {{\L}astowiecki}, \citenamefont {Beni\'{c}},
  \citenamefont {Blaschke}, \citenamefont {Maruyama},\ and\ \citenamefont
  {Tatsumi}}]{Yasutake2014_PRC89-065803}%
  \BibitemOpen
  \bibfield  {author} {\bibinfo {author} {\bibfnamefont {N.}~\bibnamefont
  {Yasutake}}, \bibinfo {author} {\bibfnamefont {R.}~\bibnamefont
  {{\L}astowiecki}}, \bibinfo {author} {\bibfnamefont {S.}~\bibnamefont
  {Beni\'{c}}}, \bibinfo {author} {\bibfnamefont {D.}~\bibnamefont {Blaschke}},
  \bibinfo {author} {\bibfnamefont {T.}~\bibnamefont {Maruyama}}, \ and\
  \bibinfo {author} {\bibfnamefont {T.}~\bibnamefont {Tatsumi}},\ }\href
  {\doibase 10.1103/PhysRevC.89.065803} {\bibfield  {journal} {\bibinfo
  {journal} {Phys. Rev. C}\ }\textbf {\bibinfo {volume} {89}},\ \bibinfo
  {pages} {065803} (\bibinfo {year} {2014})}\BibitemShut {NoStop}%
\bibitem [{\citenamefont {Xia}\ \emph {et~al.}(2019)\citenamefont {Xia},
  \citenamefont {Maruyama}, \citenamefont {Yasutake},\ and\ \citenamefont
  {Tatsumi}}]{Xia2019_PRD99-103017}%
  \BibitemOpen
  \bibfield  {author} {\bibinfo {author} {\bibfnamefont {C.-J.}\ \bibnamefont
  {Xia}}, \bibinfo {author} {\bibfnamefont {T.}~\bibnamefont {Maruyama}},
  \bibinfo {author} {\bibfnamefont {N.}~\bibnamefont {Yasutake}}, \ and\
  \bibinfo {author} {\bibfnamefont {T.}~\bibnamefont {Tatsumi}},\ }\href
  {\doibase 10.1103/PhysRevD.99.103017} {\bibfield  {journal} {\bibinfo
  {journal} {Phys. Rev. D}\ }\textbf {\bibinfo {volume} {99}},\ \bibinfo
  {pages} {103017} (\bibinfo {year} {2019})}\BibitemShut {NoStop}%
\bibitem [{\citenamefont {Maslov}\ \emph {et~al.}(2019)\citenamefont {Maslov},
  \citenamefont {Yasutake}, \citenamefont {Blaschke}, \citenamefont {Ayriyan},
  \citenamefont {Grigorian}, \citenamefont {Maruyama}, \citenamefont
  {Tatsumi},\ and\ \citenamefont {Voskresensky}}]{Maslov2019_PRC100-025802}%
  \BibitemOpen
  \bibfield  {author} {\bibinfo {author} {\bibfnamefont {K.}~\bibnamefont
  {Maslov}}, \bibinfo {author} {\bibfnamefont {N.}~\bibnamefont {Yasutake}},
  \bibinfo {author} {\bibfnamefont {D.}~\bibnamefont {Blaschke}}, \bibinfo
  {author} {\bibfnamefont {A.}~\bibnamefont {Ayriyan}}, \bibinfo {author}
  {\bibfnamefont {H.}~\bibnamefont {Grigorian}}, \bibinfo {author}
  {\bibfnamefont {T.}~\bibnamefont {Maruyama}}, \bibinfo {author}
  {\bibfnamefont {T.}~\bibnamefont {Tatsumi}}, \ and\ \bibinfo {author}
  {\bibfnamefont {D.~N.}\ \bibnamefont {Voskresensky}},\ }\href {\doibase
  10.1103/PhysRevC.100.025802} {\bibfield  {journal} {\bibinfo  {journal}
  {Phys. Rev. C}\ }\textbf {\bibinfo {volume} {100}},\ \bibinfo {pages}
  {025802} (\bibinfo {year} {2019})}\BibitemShut {NoStop}%
\bibitem [{\citenamefont {Xia}\ \emph {et~al.}(2020)\citenamefont {Xia},
  \citenamefont {Maruyama}, \citenamefont {Yasutake}, \citenamefont {Tatsumi},
  \citenamefont {Shen},\ and\ \citenamefont {Togashi}}]{Xia2020_PRD102-023031}%
  \BibitemOpen
  \bibfield  {author} {\bibinfo {author} {\bibfnamefont {C.-J.}\ \bibnamefont
  {Xia}}, \bibinfo {author} {\bibfnamefont {T.}~\bibnamefont {Maruyama}},
  \bibinfo {author} {\bibfnamefont {N.}~\bibnamefont {Yasutake}}, \bibinfo
  {author} {\bibfnamefont {T.}~\bibnamefont {Tatsumi}}, \bibinfo {author}
  {\bibfnamefont {H.}~\bibnamefont {Shen}}, \ and\ \bibinfo {author}
  {\bibfnamefont {H.}~\bibnamefont {Togashi}},\ }\href {\doibase
  10.1103/PhysRevD.102.023031} {\bibfield  {journal} {\bibinfo  {journal}
  {Phys. Rev. D}\ }\textbf {\bibinfo {volume} {102}},\ \bibinfo {pages}
  {023031} (\bibinfo {year} {2020})}\BibitemShut {NoStop}%
\bibitem [{\citenamefont {Zhang}\ \emph {et~al.}(2018)\citenamefont {Zhang},
  \citenamefont {Li},\ and\ \citenamefont {Xu}}]{Zhang2018_ApJ859-90}%
  \BibitemOpen
  \bibfield  {author} {\bibinfo {author} {\bibfnamefont {N.-B.}\ \bibnamefont
  {Zhang}}, \bibinfo {author} {\bibfnamefont {B.-A.}\ \bibnamefont {Li}}, \
  and\ \bibinfo {author} {\bibfnamefont {J.}~\bibnamefont {Xu}},\ }\href
  {\doibase 10.3847/1538-4357/aac027} {\bibfield  {journal} {\bibinfo
  {journal} {Astrophys. J.}\ }\textbf {\bibinfo {volume} {859}},\ \bibinfo
  {pages} {90} (\bibinfo {year} {2018})}\BibitemShut {NoStop}%
\bibitem [{\citenamefont {Peng}\ \emph {et~al.}(2000)\citenamefont {Peng},
  \citenamefont {Chiang}, \citenamefont {Zou}, \citenamefont {Ning},\ and\
  \citenamefont {Luo}}]{Peng2000_PRC62-025801}%
  \BibitemOpen
  \bibfield  {author} {\bibinfo {author} {\bibfnamefont {G.~X.}\ \bibnamefont
  {Peng}}, \bibinfo {author} {\bibfnamefont {H.~C.}\ \bibnamefont {Chiang}},
  \bibinfo {author} {\bibfnamefont {B.~S.}\ \bibnamefont {Zou}}, \bibinfo
  {author} {\bibfnamefont {P.~Z.}\ \bibnamefont {Ning}}, \ and\ \bibinfo
  {author} {\bibfnamefont {S.~J.}\ \bibnamefont {Luo}},\ }\href {\doibase
  10.1103/PhysRevC.62.025801} {\bibfield  {journal} {\bibinfo  {journal} {Phys.
  Rev. C}\ }\textbf {\bibinfo {volume} {62}},\ \bibinfo {pages} {025801}
  (\bibinfo {year} {2000})}\BibitemShut {NoStop}%
\bibitem [{\citenamefont {Wen}\ \emph {et~al.}(2005)\citenamefont {Wen},
  \citenamefont {Zhong}, \citenamefont {Peng}, \citenamefont {Shen},\ and\
  \citenamefont {Ning}}]{Wen2005_PRC72-015204}%
  \BibitemOpen
  \bibfield  {author} {\bibinfo {author} {\bibfnamefont {X.~J.}\ \bibnamefont
  {Wen}}, \bibinfo {author} {\bibfnamefont {X.~H.}\ \bibnamefont {Zhong}},
  \bibinfo {author} {\bibfnamefont {G.~X.}\ \bibnamefont {Peng}}, \bibinfo
  {author} {\bibfnamefont {P.~N.}\ \bibnamefont {Shen}}, \ and\ \bibinfo
  {author} {\bibfnamefont {P.~Z.}\ \bibnamefont {Ning}},\ }\href {\doibase
  10.1103/PhysRevC.72.015204} {\bibfield  {journal} {\bibinfo  {journal} {Phys.
  Rev. C}\ }\textbf {\bibinfo {volume} {72}},\ \bibinfo {pages} {015204}
  (\bibinfo {year} {2005})}\BibitemShut {NoStop}%
\bibitem [{\citenamefont {Xia}\ \emph {et~al.}(2018)\citenamefont {Xia},
  \citenamefont {Peng}, \citenamefont {Sun}, \citenamefont {Guo}, \citenamefont
  {Lu},\ and\ \citenamefont {Jaikumar}}]{Xia2018_PRD98-034031}%
  \BibitemOpen
  \bibfield  {author} {\bibinfo {author} {\bibfnamefont {C.-J.}\ \bibnamefont
  {Xia}}, \bibinfo {author} {\bibfnamefont {G.-X.}\ \bibnamefont {Peng}},
  \bibinfo {author} {\bibfnamefont {T.-T.}\ \bibnamefont {Sun}}, \bibinfo
  {author} {\bibfnamefont {W.-L.}\ \bibnamefont {Guo}}, \bibinfo {author}
  {\bibfnamefont {D.-H.}\ \bibnamefont {Lu}}, \ and\ \bibinfo {author}
  {\bibfnamefont {P.}~\bibnamefont {Jaikumar}},\ }\href {\doibase
  10.1103/PhysRevD.98.034031} {\bibfield  {journal} {\bibinfo  {journal} {Phys.
  Rev. D}\ }\textbf {\bibinfo {volume} {98}},\ \bibinfo {pages} {034031}
  (\bibinfo {year} {2018})}\BibitemShut {NoStop}%
\bibitem [{\citenamefont {Xia}(2019)}]{Xia2019_AIPCP2127-020029}%
  \BibitemOpen
  \bibfield  {author} {\bibinfo {author} {\bibfnamefont {C.-J.}\ \bibnamefont
  {Xia}},\ }\href {\doibase 10.1063/1.5117819} {\bibfield  {journal} {\bibinfo
  {journal} {AIP Conf. Proc.}\ }\textbf {\bibinfo {volume} {2127}},\ \bibinfo
  {pages} {020029} (\bibinfo {year} {2019})}\BibitemShut {NoStop}%
\bibitem [{\citenamefont {Pfaff}\ \emph {et~al.}(2022)\citenamefont {Pfaff},
  \citenamefont {Hansen},\ and\ \citenamefont {Gulminelli}}]{Pfaff2022}%
  \BibitemOpen
  \bibfield  {author} {\bibinfo {author} {\bibfnamefont {A.}~\bibnamefont
  {Pfaff}}, \bibinfo {author} {\bibfnamefont {H.}~\bibnamefont {Hansen}}, \
  and\ \bibinfo {author} {\bibfnamefont {F.}~\bibnamefont {Gulminelli}},\
  }\href {\doibase https://doi.org/10.1103/PhysRevC.105.035802} {\bibfield
  {journal} {\bibinfo  {journal} {Phys. Rev. C}\ }\textbf {\bibinfo {volume}
  {105}},\ \bibinfo {pages} {035802} (\bibinfo {year} {2022})}\BibitemShut
  {NoStop}%
\bibitem [{\citenamefont {Peng}\ \emph {et~al.}(2002)\citenamefont {Peng},
  \citenamefont {Lombardo}, \citenamefont {Loewe},\ and\ \citenamefont
  {Chiang}}]{Peng2002_PLB548-189}%
  \BibitemOpen
  \bibfield  {author} {\bibinfo {author} {\bibfnamefont {G.~X.}\ \bibnamefont
  {Peng}}, \bibinfo {author} {\bibfnamefont {U.}~\bibnamefont {Lombardo}},
  \bibinfo {author} {\bibfnamefont {M.}~\bibnamefont {Loewe}}, \ and\ \bibinfo
  {author} {\bibfnamefont {H.~C.}\ \bibnamefont {Chiang}},\ }\href {\doibase
  10.1016/S0370-2693(02)02842-3} {\bibfield  {journal} {\bibinfo  {journal}
  {Phys. Lett. B}\ }\textbf {\bibinfo {volume} {548}},\ \bibinfo {pages} {189}
  (\bibinfo {year} {2002})}\BibitemShut {NoStop}%
\bibitem [{\citenamefont {Minamikawa}\ \emph {et~al.}(2021)\citenamefont
  {Minamikawa}, \citenamefont {Kojo},\ and\ \citenamefont
  {Harada}}]{Minamikawa2021}%
  \BibitemOpen
  \bibfield  {author} {\bibinfo {author} {\bibfnamefont {T.}~\bibnamefont
  {Minamikawa}}, \bibinfo {author} {\bibfnamefont {T.}~\bibnamefont {Kojo}}, \
  and\ \bibinfo {author} {\bibfnamefont {M.}~\bibnamefont {Harada}},\ }\href
  {\doibase https://doi.org/10.1103/PhysRevC.104.065201} {\bibfield  {journal}
  {\bibinfo  {journal} {Physical Review C}\ }\textbf {\bibinfo {volume}
  {104}},\ \bibinfo {pages} {065201} (\bibinfo {year} {2021})}\BibitemShut
  {NoStop}%
\bibitem [{\citenamefont {Jin}\ \emph {et~al.}(2022)\citenamefont {Jin},
  \citenamefont {Xia}, \citenamefont {Sun},\ and\ \citenamefont
  {Peng}}]{Jin2022}%
  \BibitemOpen
  \bibfield  {author} {\bibinfo {author} {\bibfnamefont {H.-M.}\ \bibnamefont
  {Jin}}, \bibinfo {author} {\bibfnamefont {C.-J.}\ \bibnamefont {Xia}},
  \bibinfo {author} {\bibfnamefont {T.-T.}\ \bibnamefont {Sun}}, \ and\
  \bibinfo {author} {\bibfnamefont {G.-X.}\ \bibnamefont {Peng}},\ }\href
  {https://doi.org/10.1016/j.physletb.2022.137121} {\bibfield  {journal}
  {\bibinfo  {journal} {Phys. Lett. B}\ ,\ \bibinfo {pages} {137121}} (\bibinfo
  {year} {2022})}\BibitemShut {NoStop}%
\bibitem [{\citenamefont {Shlomo}\ \emph {et~al.}(2006)\citenamefont {Shlomo},
  \citenamefont {Kolomietz},\ and\ \citenamefont
  {Col{\`o}}}]{Shlomo2006_EPJA30-23}%
  \BibitemOpen
  \bibfield  {author} {\bibinfo {author} {\bibfnamefont {S.}~\bibnamefont
  {Shlomo}}, \bibinfo {author} {\bibfnamefont {V.~M.}\ \bibnamefont
  {Kolomietz}}, \ and\ \bibinfo {author} {\bibfnamefont {G.}~\bibnamefont
  {Col{\`o}}},\ }\href {\doibase 10.1140/epja/i2006-10100-3} {\bibfield
  {journal} {\bibinfo  {journal} {Eur. Phys. J. A}\ }\textbf {\bibinfo {volume}
  {30}},\ \bibinfo {pages} {23} (\bibinfo {year} {2006})}\BibitemShut {NoStop}%
\bibitem [{\citenamefont {Li}\ and\ \citenamefont
  {Han}(2013)}]{Li2013_PLB727-276}%
  \BibitemOpen
  \bibfield  {author} {\bibinfo {author} {\bibfnamefont {B.-A.}\ \bibnamefont
  {Li}}\ and\ \bibinfo {author} {\bibfnamefont {X.}~\bibnamefont {Han}},\
  }\href {\doibase http://dx.doi.org/10.1016/j.physletb.2013.10.006} {\bibfield
   {journal} {\bibinfo  {journal} {Phys. Lett. B}\ }\textbf {\bibinfo {volume}
  {727}},\ \bibinfo {pages} {276 } (\bibinfo {year} {2013})}\BibitemShut
  {NoStop}%
\bibitem [{\citenamefont {Oertel}\ \emph {et~al.}(2017)\citenamefont {Oertel},
  \citenamefont {Hempel}, \citenamefont {Kl\"ahn},\ and\ \citenamefont
  {Typel}}]{Oertel2017_RMP89-015007}%
  \BibitemOpen
  \bibfield  {author} {\bibinfo {author} {\bibfnamefont {M.}~\bibnamefont
  {Oertel}}, \bibinfo {author} {\bibfnamefont {M.}~\bibnamefont {Hempel}},
  \bibinfo {author} {\bibfnamefont {T.}~\bibnamefont {Kl\"ahn}}, \ and\
  \bibinfo {author} {\bibfnamefont {S.}~\bibnamefont {Typel}},\ }\href
  {\doibase 10.1103/RevModPhys.89.015007} {\bibfield  {journal} {\bibinfo
  {journal} {Rev. Mod. Phys.}\ }\textbf {\bibinfo {volume} {89}},\ \bibinfo
  {pages} {015007} (\bibinfo {year} {2017})}\BibitemShut {NoStop}%
\bibitem [{\citenamefont {Zhang}\ \emph {et~al.}(2020)\citenamefont {Zhang},
  \citenamefont {Liu}, \citenamefont {Xia}, \citenamefont {Li},\ and\
  \citenamefont {Biswal}}]{Zhang2020_PRC101-034303}%
  \BibitemOpen
  \bibfield  {author} {\bibinfo {author} {\bibfnamefont {Y.}~\bibnamefont
  {Zhang}}, \bibinfo {author} {\bibfnamefont {M.}~\bibnamefont {Liu}}, \bibinfo
  {author} {\bibfnamefont {C.-J.}\ \bibnamefont {Xia}}, \bibinfo {author}
  {\bibfnamefont {Z.}~\bibnamefont {Li}}, \ and\ \bibinfo {author}
  {\bibfnamefont {S.~K.}\ \bibnamefont {Biswal}},\ }\href {\doibase
  10.1103/PhysRevC.101.034303} {\bibfield  {journal} {\bibinfo  {journal}
  {Phys. Rev. C}\ }\textbf {\bibinfo {volume} {101}},\ \bibinfo {pages}
  {034303} (\bibinfo {year} {2020})}\BibitemShut {NoStop}%
\bibitem [{\citenamefont {Essick}\ \emph {et~al.}(2021)\citenamefont {Essick},
  \citenamefont {Tews}, \citenamefont {Landry},\ and\ \citenamefont
  {Schwenk}}]{Essick2021_PRL127-192701}%
  \BibitemOpen
  \bibfield  {author} {\bibinfo {author} {\bibfnamefont {R.}~\bibnamefont
  {Essick}}, \bibinfo {author} {\bibfnamefont {I.}~\bibnamefont {Tews}},
  \bibinfo {author} {\bibfnamefont {P.}~\bibnamefont {Landry}}, \ and\ \bibinfo
  {author} {\bibfnamefont {A.}~\bibnamefont {Schwenk}},\ }\href {\doibase
  10.1103/PhysRevLett.127.192701} {\bibfield  {journal} {\bibinfo  {journal}
  {Phys. Rev. Lett.}\ }\textbf {\bibinfo {volume} {127}},\ \bibinfo {pages}
  {192701} (\bibinfo {year} {2021})}\BibitemShut {NoStop}%
\bibitem [{\citenamefont {Li}\ \emph {et~al.}(2021)\citenamefont {Li},
  \citenamefont {Cai}, \citenamefont {Xie},\ and\ \citenamefont
  {Zhang}}]{Li2021_Universe7-182}%
  \BibitemOpen
  \bibfield  {author} {\bibinfo {author} {\bibfnamefont {B.-A.}\ \bibnamefont
  {Li}}, \bibinfo {author} {\bibfnamefont {B.-J.}\ \bibnamefont {Cai}},
  \bibinfo {author} {\bibfnamefont {W.-J.}\ \bibnamefont {Xie}}, \ and\
  \bibinfo {author} {\bibfnamefont {N.-B.}\ \bibnamefont {Zhang}},\ }\href
  {\doibase 10.3390/universe7060182} {\bibfield  {journal} {\bibinfo  {journal}
  {Universe}\ }\textbf {\bibinfo {volume} {7}},\ \bibinfo {pages} {182}
  (\bibinfo {year} {2021})}\BibitemShut {NoStop}%
\bibitem [{\citenamefont {Centelles}\ \emph {et~al.}(2009)\citenamefont
  {Centelles}, \citenamefont {Roca-Maza}, \citenamefont {Vi\~nas},\ and\
  \citenamefont {Warda}}]{Centelles2009_PRL102-122502}%
  \BibitemOpen
  \bibfield  {author} {\bibinfo {author} {\bibfnamefont {M.}~\bibnamefont
  {Centelles}}, \bibinfo {author} {\bibfnamefont {X.}~\bibnamefont
  {Roca-Maza}}, \bibinfo {author} {\bibfnamefont {X.}~\bibnamefont {Vi\~nas}},
  \ and\ \bibinfo {author} {\bibfnamefont {M.}~\bibnamefont {Warda}},\ }\href
  {\doibase 10.1103/PhysRevLett.102.122502} {\bibfield  {journal} {\bibinfo
  {journal} {Phys. Rev. Lett.}\ }\textbf {\bibinfo {volume} {102}},\ \bibinfo
  {pages} {122502} (\bibinfo {year} {2009})}\BibitemShut {NoStop}%
\bibitem [{\citenamefont {Brown}(2013)}]{Brown2013_PRL111-232502}%
  \BibitemOpen
  \bibfield  {author} {\bibinfo {author} {\bibfnamefont {B.~A.}\ \bibnamefont
  {Brown}},\ }\href {\doibase 10.1103/PhysRevLett.111.232502} {\bibfield
  {journal} {\bibinfo  {journal} {Phys. Rev. Lett.}\ }\textbf {\bibinfo
  {volume} {111}},\ \bibinfo {pages} {232502} (\bibinfo {year}
  {2013})}\BibitemShut {NoStop}%
\bibitem [{\citenamefont {Horowitz}\ and\ \citenamefont
  {Piekarewicz}(2001)}]{Horowitz2001_PRL86-5647}%
  \BibitemOpen
  \bibfield  {author} {\bibinfo {author} {\bibfnamefont {C.~J.}\ \bibnamefont
  {Horowitz}}\ and\ \bibinfo {author} {\bibfnamefont {J.}~\bibnamefont
  {Piekarewicz}},\ }\href {\doibase 10.1103/PhysRevLett.86.5647} {\bibfield
  {journal} {\bibinfo  {journal} {Phys. Rev. Lett.}\ }\textbf {\bibinfo
  {volume} {86}},\ \bibinfo {pages} {5647} (\bibinfo {year}
  {2001})}\BibitemShut {NoStop}%
\bibitem [{\citenamefont {Margueron}\ \emph {et~al.}(2018)\citenamefont
  {Margueron}, \citenamefont {Hoffmann~Casali},\ and\ \citenamefont
  {Gulminelli}}]{Margueron2018_PRC97-025805}%
  \BibitemOpen
  \bibfield  {author} {\bibinfo {author} {\bibfnamefont {J.}~\bibnamefont
  {Margueron}}, \bibinfo {author} {\bibfnamefont {R.}~\bibnamefont
  {Hoffmann~Casali}}, \ and\ \bibinfo {author} {\bibfnamefont {F.}~\bibnamefont
  {Gulminelli}},\ }\href {\doibase 10.1103/PhysRevC.97.025805} {\bibfield
  {journal} {\bibinfo  {journal} {Phys. Rev. C}\ }\textbf {\bibinfo {volume}
  {97}},\ \bibinfo {pages} {025805} (\bibinfo {year} {2018})}\BibitemShut
  {NoStop}%
\bibitem [{\citenamefont {Cai}\ and\ \citenamefont
  {Li}(2021)}]{Cai2021_PRC103-054611}%
  \BibitemOpen
  \bibfield  {author} {\bibinfo {author} {\bibfnamefont {B.-J.}\ \bibnamefont
  {Cai}}\ and\ \bibinfo {author} {\bibfnamefont {B.-A.}\ \bibnamefont {Li}},\
  }\href {\doibase 10.1103/PhysRevC.103.054611} {\bibfield  {journal} {\bibinfo
   {journal} {Phys. Rev. C}\ }\textbf {\bibinfo {volume} {103}},\ \bibinfo
  {pages} {054611} (\bibinfo {year} {2021})}\BibitemShut {NoStop}%
\bibitem [{\citenamefont {Peng}\ \emph {et~al.}(1999)\citenamefont {Peng},
  \citenamefont {Chiang}, \citenamefont {Yang}, \citenamefont {Li},\ and\
  \citenamefont {Liu}}]{Peng1999_PRC61-015201}%
  \BibitemOpen
  \bibfield  {author} {\bibinfo {author} {\bibfnamefont {G.~X.}\ \bibnamefont
  {Peng}}, \bibinfo {author} {\bibfnamefont {H.~C.}\ \bibnamefont {Chiang}},
  \bibinfo {author} {\bibfnamefont {J.~J.}\ \bibnamefont {Yang}}, \bibinfo
  {author} {\bibfnamefont {L.}~\bibnamefont {Li}}, \ and\ \bibinfo {author}
  {\bibfnamefont {B.}~\bibnamefont {Liu}},\ }\href {\doibase
  10.1103/PhysRevC.61.015201} {\bibfield  {journal} {\bibinfo  {journal} {Phys.
  Rev. C}\ }\textbf {\bibinfo {volume} {61}},\ \bibinfo {pages} {015201}
  (\bibinfo {year} {1999})}\BibitemShut {NoStop}%
\bibitem [{\citenamefont {Xia}\ \emph {et~al.}(2014)\citenamefont {Xia},
  \citenamefont {Peng}, \citenamefont {Chen}, \citenamefont {Lu},\ and\
  \citenamefont {Xu}}]{Xia2014_PRD89-105027}%
  \BibitemOpen
  \bibfield  {author} {\bibinfo {author} {\bibfnamefont {C.~J.}\ \bibnamefont
  {Xia}}, \bibinfo {author} {\bibfnamefont {G.~X.}\ \bibnamefont {Peng}},
  \bibinfo {author} {\bibfnamefont {S.~W.}\ \bibnamefont {Chen}}, \bibinfo
  {author} {\bibfnamefont {Z.~Y.}\ \bibnamefont {Lu}}, \ and\ \bibinfo {author}
  {\bibfnamefont {J.~F.}\ \bibnamefont {Xu}},\ }\href {\doibase
  10.1103/PhysRevD.89.105027} {\bibfield  {journal} {\bibinfo  {journal} {Phys.
  Rev. D}\ }\textbf {\bibinfo {volume} {89}},\ \bibinfo {pages} {105027}
  (\bibinfo {year} {2014})}\BibitemShut {NoStop}%
\bibitem [{\citenamefont {Xie}\ and\ \citenamefont {Li}(2021)}]{Xie2021}%
  \BibitemOpen
  \bibfield  {author} {\bibinfo {author} {\bibfnamefont {W.-J.}\ \bibnamefont
  {Xie}}\ and\ \bibinfo {author} {\bibfnamefont {B.-A.}\ \bibnamefont {Li}},\
  }\href {\doibase https://doi.org/10.1103/PhysRevC.103.035802} {\bibfield
  {journal} {\bibinfo  {journal} {Phys. Rev. C}\ }\textbf {\bibinfo {volume}
  {103}},\ \bibinfo {pages} {035802} (\bibinfo {year} {2021})}\BibitemShut
  {NoStop}%
\bibitem [{\citenamefont {Ang}\ \emph {et~al.}(2009)\citenamefont {Ang},
  \citenamefont {Guang-Xiong},\ and\ \citenamefont {Lombardo}}]{Ang2009}%
  \BibitemOpen
  \bibfield  {author} {\bibinfo {author} {\bibfnamefont {L.}~\bibnamefont
  {Ang}}, \bibinfo {author} {\bibfnamefont {P.}~\bibnamefont {Guang-Xiong}}, \
  and\ \bibinfo {author} {\bibfnamefont {U.}~\bibnamefont {Lombardo}},\ }\href
  {\doibase 10.1088/1674-1137/33/s1/020} {\bibfield  {journal} {\bibinfo
  {journal} {Chin. Phys. C}\ }\textbf {\bibinfo {volume} {33}},\ \bibinfo
  {pages} {61} (\bibinfo {year} {2009})}\BibitemShut {NoStop}%
\bibitem [{\citenamefont {Cohen}\ \emph {et~al.}(1992)\citenamefont {Cohen},
  \citenamefont {Furnstahl},\ and\ \citenamefont
  {Griegel}}]{Cohen1992_PRC45-1881}%
  \BibitemOpen
  \bibfield  {author} {\bibinfo {author} {\bibfnamefont {T.~D.}\ \bibnamefont
  {Cohen}}, \bibinfo {author} {\bibfnamefont {R.~J.}\ \bibnamefont
  {Furnstahl}}, \ and\ \bibinfo {author} {\bibfnamefont {D.~K.}\ \bibnamefont
  {Griegel}},\ }\href {\doibase 10.1103/PhysRevC.45.1881} {\bibfield  {journal}
  {\bibinfo  {journal} {Phys. Rev. C}\ }\textbf {\bibinfo {volume} {45}},\
  \bibinfo {pages} {1881} (\bibinfo {year} {1992})}\BibitemShut {NoStop}%
\bibitem [{\citenamefont {Gell-Mann}\ \emph {et~al.}(1968)\citenamefont
  {Gell-Mann}, \citenamefont {Oakes},\ and\ \citenamefont
  {Renner}}]{Gell-Mann1968_PR175-2195}%
  \BibitemOpen
  \bibfield  {author} {\bibinfo {author} {\bibfnamefont {M.}~\bibnamefont
  {Gell-Mann}}, \bibinfo {author} {\bibfnamefont {R.~J.}\ \bibnamefont
  {Oakes}}, \ and\ \bibinfo {author} {\bibfnamefont {B.}~\bibnamefont
  {Renner}},\ }\href {\doibase 10.1103/PhysRev.175.2195} {\bibfield  {journal}
  {\bibinfo  {journal} {Phys. Rev.}\ }\textbf {\bibinfo {volume} {175}},\
  \bibinfo {pages} {2195} (\bibinfo {year} {1968})}\BibitemShut {NoStop}%
\bibitem [{\citenamefont {Abbott}\ \emph {et~al.}(2019)\citenamefont {Abbott},
  \citenamefont {Abbott}, \citenamefont {Abbott}, \citenamefont {Abraham},
  \citenamefont {Acernese}, \citenamefont {Ackley}, \citenamefont {Adams},
  \citenamefont {Adhikari}, \citenamefont {Adya}, \citenamefont {Affeldt} \emph
  {et~al.}}]{Abbott2019}%
  \BibitemOpen
  \bibfield  {author} {\bibinfo {author} {\bibfnamefont {B.~P.}\ \bibnamefont
  {Abbott}}, \bibinfo {author} {\bibfnamefont {R.}~\bibnamefont {Abbott}},
  \bibinfo {author} {\bibfnamefont {T.}~\bibnamefont {Abbott}}, \bibinfo
  {author} {\bibfnamefont {S.}~\bibnamefont {Abraham}}, \bibinfo {author}
  {\bibfnamefont {F.}~\bibnamefont {Acernese}}, \bibinfo {author}
  {\bibfnamefont {K.}~\bibnamefont {Ackley}}, \bibinfo {author} {\bibfnamefont
  {C.}~\bibnamefont {Adams}}, \bibinfo {author} {\bibfnamefont {R.~X.}\
  \bibnamefont {Adhikari}}, \bibinfo {author} {\bibfnamefont {V.}~\bibnamefont
  {Adya}}, \bibinfo {author} {\bibfnamefont {C.}~\bibnamefont {Affeldt}},
  \emph {et~al.},\ }\href {\doibase
  https://doi.org/10.1103/PhysRevLett.123.161102} {\bibfield  {journal}
  {\bibinfo  {journal} {Phys. Rev. Lett}\ }\textbf {\bibinfo {volume} {123}},\
  \bibinfo {pages} {161102} (\bibinfo {year} {2019})}\BibitemShut {NoStop}%
\bibitem [{\citenamefont {Cromartie}\ \emph {et~al.}(2020)\citenamefont
  {Cromartie}, \citenamefont {Fonseca}, \citenamefont {Ransom}, \citenamefont
  {Demorest}, \citenamefont {Arzoumanian}, \citenamefont {Blumer},
  \citenamefont {Brook}, \citenamefont {DeCesar}, \citenamefont {Dolch},
  \citenamefont {Ellis}, \citenamefont {Ferdman}, \citenamefont {Ferrara},
  \citenamefont {Garver-Daniels}, \citenamefont {Gentile}, \citenamefont
  {Jones}, \citenamefont {Lam}, \citenamefont {Lorimer}, \citenamefont {Lynch},
  \citenamefont {McLaughlin}, \citenamefont {Ng}, \citenamefont {Nice},
  \citenamefont {Pennucci}, \citenamefont {Spiewak}, \citenamefont {Stairs},
  \citenamefont {Stovall}, \citenamefont {Swiggum},\ and\ \citenamefont
  {Zhu}}]{Cromartie2020_NA4-72}%
  \BibitemOpen
  \bibfield  {author} {\bibinfo {author} {\bibfnamefont {H.~T.}\ \bibnamefont
  {Cromartie}}, \bibinfo {author} {\bibfnamefont {E.}~\bibnamefont {Fonseca}},
  \bibinfo {author} {\bibfnamefont {S.~M.}\ \bibnamefont {Ransom}}, \bibinfo
  {author} {\bibfnamefont {P.~B.}\ \bibnamefont {Demorest}}, \bibinfo {author}
  {\bibfnamefont {Z.}~\bibnamefont {Arzoumanian}}, \bibinfo {author}
  {\bibfnamefont {H.}~\bibnamefont {Blumer}}, \bibinfo {author} {\bibfnamefont
  {P.~R.}\ \bibnamefont {Brook}}, \bibinfo {author} {\bibfnamefont {M.~E.}\
  \bibnamefont {DeCesar}}, \bibinfo {author} {\bibfnamefont {T.}~\bibnamefont
  {Dolch}}, \bibinfo {author} {\bibfnamefont {J.~A.}\ \bibnamefont {Ellis}},
  \bibinfo {author} {\bibfnamefont {R.~D.}\ \bibnamefont {Ferdman}}, \bibinfo
  {author} {\bibfnamefont {E.~C.}\ \bibnamefont {Ferrara}}, \bibinfo {author}
  {\bibfnamefont {N.}~\bibnamefont {Garver-Daniels}}, \bibinfo {author}
  {\bibfnamefont {P.~A.}\ \bibnamefont {Gentile}}, \bibinfo {author}
  {\bibfnamefont {M.~L.}\ \bibnamefont {Jones}}, \bibinfo {author}
  {\bibfnamefont {M.~T.}\ \bibnamefont {Lam}}, \bibinfo {author} {\bibfnamefont
  {D.~R.}\ \bibnamefont {Lorimer}}, \bibinfo {author} {\bibfnamefont {R.~S.}\
  \bibnamefont {Lynch}}, \bibinfo {author} {\bibfnamefont {M.~A.}\ \bibnamefont
  {McLaughlin}}, \bibinfo {author} {\bibfnamefont {C.}~\bibnamefont {Ng}},
  \bibinfo {author} {\bibfnamefont {D.~J.}\ \bibnamefont {Nice}}, \bibinfo
  {author} {\bibfnamefont {T.~T.}\ \bibnamefont {Pennucci}}, \bibinfo {author}
  {\bibfnamefont {R.}~\bibnamefont {Spiewak}}, \bibinfo {author} {\bibfnamefont
  {I.~H.}\ \bibnamefont {Stairs}}, \bibinfo {author} {\bibfnamefont
  {K.}~\bibnamefont {Stovall}}, \bibinfo {author} {\bibfnamefont {J.~K.}\
  \bibnamefont {Swiggum}}, \ and\ \bibinfo {author} {\bibfnamefont
  {W.}~\bibnamefont {Zhu}},\ }\href {\doibase 10.1038/s41550-019-0880-2}
  {\bibfield  {journal} {\bibinfo  {journal} {Nat. Astron.}\ }\textbf {\bibinfo
  {volume} {4}},\ \bibinfo {pages} {72} (\bibinfo {year} {2020})}\BibitemShut
  {NoStop}%
\bibitem [{\citenamefont {Watts}\ \emph {et~al.}(2018)\citenamefont {Watts},
  \citenamefont {Yu}, \citenamefont {Poutanen}, \citenamefont {Zhang},
  \citenamefont {Bhattacharyya}, \citenamefont {Bogdanov}, \citenamefont {Ji},
  \citenamefont {Patruno}, \citenamefont {Riley}, \citenamefont {Bakala},
  \citenamefont {Baykal}, \citenamefont {Bernardini}, \citenamefont {Bombaci},
  \citenamefont {Brown}, \citenamefont {Cavecchi}, \citenamefont {Chakrabarty},
  \citenamefont {Chenevez}, \citenamefont {Degenaar}, \citenamefont
  {Del~Santo}, \citenamefont {Di~Salvo}, \citenamefont {Doroshenko},
  \citenamefont {Falanga}, \citenamefont {Ferdman}, \citenamefont {Feroci},
  \citenamefont {Gambino}, \citenamefont {Ge}, \citenamefont {Greif},
  \citenamefont {Guillot}, \citenamefont {Gungor}, \citenamefont {Hartmann},
  \citenamefont {Hebeler}, \citenamefont {Heger}, \citenamefont {Homan},
  \citenamefont {Iaria}, \citenamefont {Zand}, \citenamefont {Kargaltsev},
  \citenamefont {Kurkela}, \citenamefont {Lai}, \citenamefont {Li},
  \citenamefont {Li}, \citenamefont {Li}, \citenamefont {Linares},
  \citenamefont {Lu}, \citenamefont {Mahmoodifar}, \citenamefont {M{\'e}ndez},
  \citenamefont {Coleman~Miller}, \citenamefont {Morsink}, \citenamefont
  {N{\"a}ttil{\"a}}, \citenamefont {Possenti}, \citenamefont
  {Prescod-Weinstein}, \citenamefont {Qu}, \citenamefont {Riggio},
  \citenamefont {Salmi}, \citenamefont {Sanna}, \citenamefont {Santangelo},
  \citenamefont {Schatz}, \citenamefont {Schwenk}, \citenamefont {Song},
  \citenamefont {{\v{S}}r{\'a}mkov{\'a}}, \citenamefont {Stappers},
  \citenamefont {Stiele}, \citenamefont {Strohmayer}, \citenamefont {Tews},
  \citenamefont {Tolos}, \citenamefont {T{\"o}r{\"o}k}, \citenamefont {Tsang},
  \citenamefont {Urbanec}, \citenamefont {Vacchi}, \citenamefont {Xu},
  \citenamefont {Xu}, \citenamefont {Zane}, \citenamefont {Zhang},
  \citenamefont {Zhang}, \citenamefont {Zhang}, \citenamefont {Zheng},\ and\
  \citenamefont {Zhou}}]{Watts2019_SCPMA62-29503}%
  \BibitemOpen
  \bibfield  {author} {\bibinfo {author} {\bibfnamefont {A.~L.}\ \bibnamefont
  {Watts}}, \bibinfo {author} {\bibfnamefont {W.}~\bibnamefont {Yu}}, \bibinfo
  {author} {\bibfnamefont {J.}~\bibnamefont {Poutanen}}, \bibinfo {author}
  {\bibfnamefont {S.}~\bibnamefont {Zhang}}, \bibinfo {author} {\bibfnamefont
  {S.}~\bibnamefont {Bhattacharyya}}, \bibinfo {author} {\bibfnamefont
  {S.}~\bibnamefont {Bogdanov}}, \bibinfo {author} {\bibfnamefont
  {L.}~\bibnamefont {Ji}}, \bibinfo {author} {\bibfnamefont {A.}~\bibnamefont
  {Patruno}}, \bibinfo {author} {\bibfnamefont {T.~E.}\ \bibnamefont {Riley}},
  \bibinfo {author} {\bibfnamefont {P.}~\bibnamefont {Bakala}}, \bibinfo
  {author} {\bibfnamefont {A.}~\bibnamefont {Baykal}}, \bibinfo {author}
  {\bibfnamefont {F.}~\bibnamefont {Bernardini}}, \bibinfo {author}
  {\bibfnamefont {I.}~\bibnamefont {Bombaci}}, \bibinfo {author} {\bibfnamefont
  {E.}~\bibnamefont {Brown}}, \bibinfo {author} {\bibfnamefont
  {Y.}~\bibnamefont {Cavecchi}}, \bibinfo {author} {\bibfnamefont
  {D.}~\bibnamefont {Chakrabarty}}, \bibinfo {author} {\bibfnamefont
  {J.}~\bibnamefont {Chenevez}}, \bibinfo {author} {\bibfnamefont
  {N.}~\bibnamefont {Degenaar}}, \bibinfo {author} {\bibfnamefont
  {M.}~\bibnamefont {Del~Santo}}, \bibinfo {author} {\bibfnamefont
  {T.}~\bibnamefont {Di~Salvo}}, \bibinfo {author} {\bibfnamefont
  {V.}~\bibnamefont {Doroshenko}}, \bibinfo {author} {\bibfnamefont
  {M.}~\bibnamefont {Falanga}}, \bibinfo {author} {\bibfnamefont {R.~D.}\
  \bibnamefont {Ferdman}}, \bibinfo {author} {\bibfnamefont {M.}~\bibnamefont
  {Feroci}}, \bibinfo {author} {\bibfnamefont {A.~F.}\ \bibnamefont {Gambino}},
  \bibinfo {author} {\bibfnamefont {M.}~\bibnamefont {Ge}}, \bibinfo {author}
  {\bibfnamefont {S.~K.}\ \bibnamefont {Greif}}, \bibinfo {author}
  {\bibfnamefont {S.}~\bibnamefont {Guillot}}, \bibinfo {author} {\bibfnamefont
  {C.}~\bibnamefont {Gungor}}, \bibinfo {author} {\bibfnamefont {D.~H.}\
  \bibnamefont {Hartmann}}, \bibinfo {author} {\bibfnamefont {K.}~\bibnamefont
  {Hebeler}}, \bibinfo {author} {\bibfnamefont {A.}~\bibnamefont {Heger}},
  \bibinfo {author} {\bibfnamefont {J.}~\bibnamefont {Homan}}, \bibinfo
  {author} {\bibfnamefont {R.}~\bibnamefont {Iaria}}, \bibinfo {author}
  {\bibfnamefont {J.~i.}\ \bibnamefont {Zand}}, \bibinfo {author}
  {\bibfnamefont {O.}~\bibnamefont {Kargaltsev}}, \bibinfo {author}
  {\bibfnamefont {A.}~\bibnamefont {Kurkela}}, \bibinfo {author} {\bibfnamefont
  {X.}~\bibnamefont {Lai}}, \bibinfo {author} {\bibfnamefont {A.}~\bibnamefont
  {Li}}, \bibinfo {author} {\bibfnamefont {X.}~\bibnamefont {Li}}, \bibinfo
  {author} {\bibfnamefont {Z.}~\bibnamefont {Li}}, \bibinfo {author}
  {\bibfnamefont {M.}~\bibnamefont {Linares}}, \bibinfo {author} {\bibfnamefont
  {F.}~\bibnamefont {Lu}}, \bibinfo {author} {\bibfnamefont {S.}~\bibnamefont
  {Mahmoodifar}}, \bibinfo {author} {\bibfnamefont {M.}~\bibnamefont
  {M{\'e}ndez}}, \bibinfo {author} {\bibfnamefont {M.}~\bibnamefont
  {Coleman~Miller}}, \bibinfo {author} {\bibfnamefont {S.}~\bibnamefont
  {Morsink}}, \bibinfo {author} {\bibfnamefont {J.}~\bibnamefont
  {N{\"a}ttil{\"a}}}, \bibinfo {author} {\bibfnamefont {A.}~\bibnamefont
  {Possenti}}, \bibinfo {author} {\bibfnamefont {C.}~\bibnamefont
  {Prescod-Weinstein}}, \bibinfo {author} {\bibfnamefont {J.}~\bibnamefont
  {Qu}}, \bibinfo {author} {\bibfnamefont {A.}~\bibnamefont {Riggio}}, \bibinfo
  {author} {\bibfnamefont {T.}~\bibnamefont {Salmi}}, \bibinfo {author}
  {\bibfnamefont {A.}~\bibnamefont {Sanna}}, \bibinfo {author} {\bibfnamefont
  {A.}~\bibnamefont {Santangelo}}, \bibinfo {author} {\bibfnamefont
  {H.}~\bibnamefont {Schatz}}, \bibinfo {author} {\bibfnamefont
  {A.}~\bibnamefont {Schwenk}}, \bibinfo {author} {\bibfnamefont
  {L.}~\bibnamefont {Song}}, \bibinfo {author} {\bibfnamefont {E.}~\bibnamefont
  {{\v{S}}r{\'a}mkov{\'a}}}, \bibinfo {author} {\bibfnamefont {B.}~\bibnamefont
  {Stappers}}, \bibinfo {author} {\bibfnamefont {H.}~\bibnamefont {Stiele}},
  \bibinfo {author} {\bibfnamefont {T.}~\bibnamefont {Strohmayer}}, \bibinfo
  {author} {\bibfnamefont {I.}~\bibnamefont {Tews}}, \bibinfo {author}
  {\bibfnamefont {L.}~\bibnamefont {Tolos}}, \bibinfo {author} {\bibfnamefont
  {G.}~\bibnamefont {T{\"o}r{\"o}k}}, \bibinfo {author} {\bibfnamefont
  {D.}~\bibnamefont {Tsang}}, \bibinfo {author} {\bibfnamefont
  {M.}~\bibnamefont {Urbanec}}, \bibinfo {author} {\bibfnamefont
  {A.}~\bibnamefont {Vacchi}}, \bibinfo {author} {\bibfnamefont
  {R.}~\bibnamefont {Xu}}, \bibinfo {author} {\bibfnamefont {Y.}~\bibnamefont
  {Xu}}, \bibinfo {author} {\bibfnamefont {S.}~\bibnamefont {Zane}}, \bibinfo
  {author} {\bibfnamefont {G.}~\bibnamefont {Zhang}}, \bibinfo {author}
  {\bibfnamefont {S.}~\bibnamefont {Zhang}}, \bibinfo {author} {\bibfnamefont
  {W.}~\bibnamefont {Zhang}}, \bibinfo {author} {\bibfnamefont
  {S.}~\bibnamefont {Zheng}}, \ and\ \bibinfo {author} {\bibfnamefont
  {X.}~\bibnamefont {Zhou}},\ }\href {\doibase 10.1007/s11433-017-9188-4}
  {\bibfield  {journal} {\bibinfo  {journal} {Sci. China Phys. Mech. Astron.}\
  }\textbf {\bibinfo {volume} {62}},\ \bibinfo {pages} {29503} (\bibinfo {year}
  {2018})}\BibitemShut {NoStop}%
\bibitem [{\citenamefont {Li}\ \emph {et~al.}(2020)\citenamefont {Li},
  \citenamefont {Zhu}, \citenamefont {Zhou}, \citenamefont {Dong},
  \citenamefont {Hu},\ and\ \citenamefont {Xia}}]{Li2020}%
  \BibitemOpen
  \bibfield  {author} {\bibinfo {author} {\bibfnamefont {A.}~\bibnamefont
  {Li}}, \bibinfo {author} {\bibfnamefont {Z.-Y.}\ \bibnamefont {Zhu}},
  \bibinfo {author} {\bibfnamefont {E.-P.}\ \bibnamefont {Zhou}}, \bibinfo
  {author} {\bibfnamefont {J.-M.}\ \bibnamefont {Dong}}, \bibinfo {author}
  {\bibfnamefont {J.-N.}\ \bibnamefont {Hu}}, \ and\ \bibinfo {author}
  {\bibfnamefont {C.-J.}\ \bibnamefont {Xia}},\ }\href {\doibase
  10.1016/j.jheap.2020.07.001} {\bibfield  {journal} {\bibinfo  {journal}
  {JHEAP}\ }\textbf {\bibinfo {volume} {28}},\ \bibinfo {pages} {19} (\bibinfo
  {year} {2020})}\BibitemShut {NoStop}%
\bibitem [{\citenamefont {Fonseca}\ \emph {et~al.}(2021)\citenamefont
  {Fonseca}, \citenamefont {Cromartie}, \citenamefont {Pennucci}, \citenamefont
  {Ray}, \citenamefont {Kirichenko}, \citenamefont {Ransom}, \citenamefont
  {Demorest}, \citenamefont {Stairs}, \citenamefont {Arzoumanian},
  \citenamefont {Guillemot} \emph {et~al.}}]{Fonseca2021}%
  \BibitemOpen
  \bibfield  {author} {\bibinfo {author} {\bibfnamefont {E.}~\bibnamefont
  {Fonseca}}, \bibinfo {author} {\bibfnamefont {H.~T.}\ \bibnamefont
  {Cromartie}}, \bibinfo {author} {\bibfnamefont {T.~T.}\ \bibnamefont
  {Pennucci}}, \bibinfo {author} {\bibfnamefont {P.~S.}\ \bibnamefont {Ray}},
  \bibinfo {author} {\bibfnamefont {A.~Y.}\ \bibnamefont {Kirichenko}},
  \bibinfo {author} {\bibfnamefont {S.~M.}\ \bibnamefont {Ransom}}, \bibinfo
  {author} {\bibfnamefont {P.~B.}\ \bibnamefont {Demorest}}, \bibinfo {author}
  {\bibfnamefont {I.~H.}\ \bibnamefont {Stairs}}, \bibinfo {author}
  {\bibfnamefont {Z.}~\bibnamefont {Arzoumanian}}, \bibinfo {author}
  {\bibfnamefont {L.}~\bibnamefont {Guillemot}},  \emph {et~al.},\ }\href
  {\doibase 10.3847/2041-8213/ac03b8} {\bibfield  {journal} {\bibinfo
  {journal} {Astrophys. J.}\ }\textbf {\bibinfo {volume} {915}},\ \bibinfo
  {pages} {L12} (\bibinfo {year} {2021})}\BibitemShut {NoStop}%
\bibitem [{\citenamefont {Damour}\ and\ \citenamefont
  {Nagar}(2009)}]{Damour2009_PRD80-084035}%
  \BibitemOpen
  \bibfield  {author} {\bibinfo {author} {\bibfnamefont {T.}~\bibnamefont
  {Damour}}\ and\ \bibinfo {author} {\bibfnamefont {A.}~\bibnamefont {Nagar}},\
  }\href {\doibase 10.1103/PhysRevD.80.084035} {\bibfield  {journal} {\bibinfo
  {journal} {Phys. Rev. D}\ }\textbf {\bibinfo {volume} {80}},\ \bibinfo
  {pages} {084035} (\bibinfo {year} {2009})}\BibitemShut {NoStop}%
\bibitem [{\citenamefont {Hinderer}\ \emph {et~al.}(2010)\citenamefont
  {Hinderer}, \citenamefont {Lackey}, \citenamefont {Lang},\ and\ \citenamefont
  {Read}}]{Hinderer2010_PRD81-123016}%
  \BibitemOpen
  \bibfield  {author} {\bibinfo {author} {\bibfnamefont {T.}~\bibnamefont
  {Hinderer}}, \bibinfo {author} {\bibfnamefont {B.~D.}\ \bibnamefont
  {Lackey}}, \bibinfo {author} {\bibfnamefont {R.~N.}\ \bibnamefont {Lang}}, \
  and\ \bibinfo {author} {\bibfnamefont {J.~S.}\ \bibnamefont {Read}},\ }\href
  {\doibase 10.1103/PhysRevD.81.123016} {\bibfield  {journal} {\bibinfo
  {journal} {Phys. Rev. D}\ }\textbf {\bibinfo {volume} {81}},\ \bibinfo
  {pages} {123016} (\bibinfo {year} {2010})}\BibitemShut {NoStop}%
\bibitem [{\citenamefont {Postnikov}\ \emph {et~al.}(2010)\citenamefont
  {Postnikov}, \citenamefont {Prakash},\ and\ \citenamefont
  {Lattimer}}]{Postnikov2010_PRD82-024016}%
  \BibitemOpen
  \bibfield  {author} {\bibinfo {author} {\bibfnamefont {S.}~\bibnamefont
  {Postnikov}}, \bibinfo {author} {\bibfnamefont {M.}~\bibnamefont {Prakash}},
  \ and\ \bibinfo {author} {\bibfnamefont {J.~M.}\ \bibnamefont {Lattimer}},\
  }\href {\doibase 10.1103/PhysRevD.82.024016} {\bibfield  {journal} {\bibinfo
  {journal} {Phys. Rev. D}\ }\textbf {\bibinfo {volume} {82}},\ \bibinfo
  {pages} {024016} (\bibinfo {year} {2010})}\BibitemShut {NoStop}%
\bibitem [{\citenamefont {Annala}\ \emph
  {et~al.}(2020{\natexlab{b}})\citenamefont {Annala}, \citenamefont {Gorda},
  \citenamefont {Kurkela}, \citenamefont {N\"attil\"a},\ and\ \citenamefont
  {Vuorinen}}]{Annala2020_NP}%
  \BibitemOpen
  \bibfield  {author} {\bibinfo {author} {\bibfnamefont {E.}~\bibnamefont
  {Annala}}, \bibinfo {author} {\bibfnamefont {T.}~\bibnamefont {Gorda}},
  \bibinfo {author} {\bibfnamefont {A.}~\bibnamefont {Kurkela}}, \bibinfo
  {author} {\bibfnamefont {J.}~\bibnamefont {N\"attil\"a}}, \ and\ \bibinfo
  {author} {\bibfnamefont {A.}~\bibnamefont {Vuorinen}},\ }\href {\doibase
  10.1038/s41567-020-0914-9} {\bibfield  {journal} {\bibinfo  {journal} {Nat.
  Phys.}\ }\textbf {\bibinfo {volume} {16}},\ \bibinfo {pages} {907} (\bibinfo
  {year} {2020}{\natexlab{b}})}\BibitemShut {NoStop}%
\bibitem [{\citenamefont {Xia}\ \emph {et~al.}(2021)\citenamefont {Xia},
  \citenamefont {Zhu}, \citenamefont {xia zhou},\ and\ \citenamefont
  {Li}}]{Xia2021_CPC45-055104}%
  \BibitemOpen
  \bibfield  {author} {\bibinfo {author} {\bibfnamefont {C.-J.}\ \bibnamefont
  {Xia}}, \bibinfo {author} {\bibfnamefont {Z.}~\bibnamefont {Zhu}}, \bibinfo
  {author} {\bibnamefont {xia zhou}}, \ and\ \bibinfo {author} {\bibfnamefont
  {A.}~\bibnamefont {Li}},\ }\href
  {http://iopscience.iop.org/article/10.1088/1674-1137/abea0d} {\bibfield
  {journal} {\bibinfo  {journal} {Chin. Phys. C}\ }\textbf {\bibinfo {volume}
  {45}},\ \bibinfo {pages} {055104} (\bibinfo {year} {2021})}\BibitemShut
  {NoStop}%
\end{thebibliography}

%

\end{document}